\documentclass[aps,prd,preprintnumbers,groupedaddress,nofootinbib,11pt]{revtex4-1}
\pdfoutput=1
\usepackage{anyfontsize}
\usepackage{graphicx}
\usepackage{latexsym}
\usepackage{amsfonts}
\usepackage{amssymb}
\usepackage{amsmath}
\usepackage{slashed}
\usepackage{array}
\usepackage{feynmp}
 \usepackage{hyperref}
\usepackage{url}
\allowdisplaybreaks

\begin{document}

%%%%%%%%%%%%%%%%%%%%%%%%%
\title{Light Dark Matter, Naturalness,
\texorpdfstring{\\}{} 
and the Radiative 
Origin of the 
Electroweak Scale}
%%%%%%%%%%%%%%%%%%%%%%%%%

\author{Wolfgang Altmannshofer$^1$, William A.~Bardeen$^2$, 
Martin~Bauer$^{2,3}$, 
Marcela~Carena$^{2,3,4}$, Joseph D.~Lykken$^2$}
\affiliation{$^1$Perimeter Institute for Theoretical Physics, Waterloo, ON, N2L 
2Y5, Canada}
\affiliation{$^2$Theoretical Physics Department, Fermi National Accelerator 
Laboratory, Batavia, IL 60510, USA}
\affiliation{$^3$Enrico Fermi Institute, University of Chicago, Chicago, IL 
60637, USA}
\affiliation{$^4$Kavli Institute for Cosmological Physics, University of 
Chicago, Chicago, IL 60637, USA}

%%%%%%%%%%%%%%%%%%%%%%%%%

\preprint{FERMILAB-PUB-14-272-T}
\preprint{EFI-14-26}

%\date{\today}

%%%%%%%%%%%%%%%%%%%%%%%%%
\begin{abstract}

We study classically scale invariant models in which the Standard Model
Higgs mass term is replaced in the Lagrangian by a Higgs portal coupling to a 
complex scalar 
field of a dark sector.
We focus on models that are weakly coupled with
the quartic scalar couplings nearly vanishing at the Planck scale.
The dark sector contains fermions and scalars charged under dark $SU(2)\times 
U(1)$ gauge interactions.
Radiative breaking of the dark gauge group triggers electroweak symmetry 
breaking
through the Higgs portal coupling.
Requiring both a Higgs boson mass of 125.5 GeV and stability of
the Higgs potential up to the Planck scale implies that the
radiative breaking of the dark gauge group occurs at the TeV scale.
We present a particular model which features a long-range abelian 
dark force. The dominant dark matter component is neutral dark fermions, with 
the correct thermal relic 
abundance, and in reach of future direct detection experiments.
The model also has lighter stable dark fermions charged under the dark 
force, with observable effects on galactic-scale structure. 
Collider signatures include a dark sector scalar boson with mass $\lesssim 250$ 
GeV 
that decays through mixing 
with the Higgs boson, and can be detected at the LHC.
The Higgs boson, as well as the new scalar, 
may have significant invisible decays into dark sector particles.

\end{abstract}
%%%%%%%%%%%%%%%%%%%%%%%%%

\maketitle

%%%%%%%%%%%%%%%%%%%%%%%%%
\section{Introduction \label{sec:intro}}

The Standard Model (SM) is a renormalizable quantum field theory that makes 
unambiguous predictions
for elementary particle processes over a very large range of energy scales. 
Apart from a possible metastable vacuum, the SM has no theoretical 
inconsistencies at least up to the Planck scale at which we expect gravity to 
become strong and quantum field theories to break down. 
If this scenario is realized in nature, the Higgs mass parameter seems 
artificially small compared to the Planck scale. However, in the SM itself the
Higgs mass parameter is the only explicit scale in the theory, and therefore
it is only multiplicatively renormalized~\cite{Bardeen:1995kv}.  

An interesting modification of the SM is given by requiring that the
Higgs mass term vanishes at some very high energy (UV) scale; in this case it 
will not be generated by 
SM radiative corrections at lower scales either. The tree-level potential has 
only a quartic term, and the full Lagrangian is classically scale invariant.
Electroweak symmetry breaking could be triggered, in principle, by 
the one-loop corrections to the effective potential
\begin{align}\label{eq:veff1}
 V_\mathrm{eff}(h)=\frac{\lambda}{2} h^4+B\,h^4\,\log(h^2/\mu^2)\,,
\end{align}
in which $\mu$ denotes the renormalization scale and $B$ is a loop suppressed 
function of the couplings. Such a possibility has been envisioned by Coleman 
and Weinberg~\cite{Coleman:1973jx}. 
A very attractive feature of the Coleman-Weinberg (CW) 
symmetry breaking mechanism is, that for couplings of order 1 at some 
renormalization scale in the UV, $\mu=\mu_\mathrm{UV}$, the minimum of the
potential appears at an exponentially smaller scale
\begin{align}
 \langle h \rangle \propto \mu_\mathrm{UV}\, 
e^{-\lambda(\mu_\mathrm{UV})/B}\,.
\end{align}
Therefore, similar to the large disparity between the Planck scale 
and the confinement scale of QCD, the large disparity between 
the Planck scale and the electroweak scale is explained through renormalization 
group running~\cite{Gildener:1976ih}.

However, in the SM the CW mechanism is ruled out. The dominant contribution to 
the effective potential comes from the top quark, which renders it 
unbounded from below, since it enters the coefficient $B$ in~\eqref{eq:veff1} 
with a negative sign. In 
order to overcome the top quark contribution and to reproduce the measured 
Higgs 
mass, one would need to extend the SM by bosonic degrees of freedom with 
sizable couplings to the Higgs \cite{Dermisek:2013pta,Hill:2014mqa}. 

Another motivation for extending the SM is the strong observational evidence 
for dark matter (DM), 
plausibly in the form of weakly interacting heavy particles. 
Even in the absence of a Higgs mass parameter in the UV,
such particles will generically
introduce additive corrections to the Higgs mass parameter and spoil the 
CW dynamics in the absence of additional symmetries.  
This motivates an alternative implementation of the CW mechanism, first 
proposed by Hempfling~\cite{Hempfling:1996ht}. In this model, the Higgs couples 
to one extra scalar, which through dynamics of a hidden sector undergoes CW 
symmetry breaking and communicates the corresponding mass scale through the 
Higgs portal to the SM. Dark matter can then be given by any of the new hidden 
sector fields that govern the renormalization group evolution of the scalar 
potential in the dark sector. 

There has been 
a lot of recent interest in models that implement various aspects of
these basic ideas \cite{Foot:2007iy,Iso:2009ss,Foot:2010av,
AlexanderNunneley:2010nw,Iso:2012jn,Englert:2013gz,Farina:2013mla,
Heikinheimo:2013fta,
Hambye:2013dgv,Carone:2013wla,Farzinnia:2013pga,Dermisek:2013pta,Khoze:2013uia,
Tamarit:2013vda,Gabrielli:2013hma,Steele:2013fka,Hashimoto:2013hta,
Holthausen:2013ota,
Hashimoto:2014ela,Hill:2014mqa,Radovcic:2014rea,Khoze:2014xha,Farzinnia:2014xia,
Pelaggi:2014wba}. Here we will focus on implementations with
dark sectors that are fairly simple and thus predictive.
In Section~\ref{sec:mot} we comment on issues
of naturalness as applied to classically scale invariant modificiations
of the SM, without claiming to resolve these issues.
In Section~\ref{sec:TeVDM} we show, that in extensions of the SM with no 
explicit mass scales, the combination of a Higgs mass term generated through CW 
symmetry breaking together with the restriction to have a stable vacuum up to 
the Planck scale generically sets an upper bound on the dark matter mass scale 
of the order 
of a few TeV. Furthermore,
the CW mechanism requires sizable couplings for gauge fields in the hidden 
sector, so that the simplest models in the literature are in addition subject 
to a lower bound on the DM mass of several hundred GeV.
In Section \ref{sec:model} we present a model with additional fermions in 
the hidden sector that can be dark matter candidates with masses  
at the 
electroweak scale or below. 
In Sections~\ref{sec:higgs},~\ref{sec:DM} and~\ref{sec:out}, we discuss the 
collider and dark matter phenomenology of the model.
In Section~\ref{sec:out}, we also comment
on further implications of this model for the dynamics of galaxy structure 
formation and a possible first order electroweak phase transition.
We conclude in Section~\ref{sec:con}.

The one loop effective potential of the discussed model 
and the one loop beta functions of the dark sector couplings
are collected in Appendices~\ref{sec:Veff} and~\ref{sec:betafunctions}. 
For the beta functions and anomalous dimensions, we follow the methods, 
conventions and notation of Machacek and 
Vaughn \cite{Machacek:1983tz,Machacek:1983fi,Machacek:1984zw},
with the improvements and extensions introduced by Luo and Xiao 
\cite{Luo:2002ey,Luo:2002ti,Luo:2002iq}.
For the effective potentials, we follow the methods and conventions of Martin 
\cite{Martin:2001vx}.
There are slight differences of notation in the literature: for example 
compared to \cite{Degrassi:2012ry,Buttazzo:2013uya},
our scalar self-coupling is twice as large, and our convention for anomalous 
dimensions has the 
opposite sign. 

%%%%%%%%%%%%%%%%%%%%%%%%%%%%%%%%%%%%%%%%%%%%%%%%%%%%%%%%%%%%%%%%
\section{Motivation}\label{sec:mot}

A Coleman-Weinberg mechanism as the origin of electroweak symmetry breaking was
first considered by Gildener and Weinberg~\cite{Gildener:1976ih}. 
In the absence of the Higgs mass term, the Lagrangian of the 
SM exhibits classical scale invariance that is softly broken by quantum 
effects -
the well known scale anomaly.
In UV completions of the 
SM, the physical thresholds associated with new massive states 
would constitute an explicit breaking of this symmetry. This introduces the 
need 
for a fine-tuning of the bare Higgs mass parameter against radiative 
corrections involving more massive particles. The fact that the Higgs mass 
parameter 
is not protected by a symmetry from these radiative corrections is known as the 
naturalness or hierarchy problem.

If the SM is UV completed by a conformal or supersymmetric (SUSY) theory, 
the Higgs mass parameter is radiatively stable above the scale 
at which this completion sets in; thus if this scale is not too high, the 
hierarchy problem is solved. This has led to the expectation that such a UV 
completion
is realized in the vicinity of the electroweak scale. 
However, the new degrees of freedom predicted by either supersymmetric or 
conformal UV completions have not been observed, yet.
This raises the prospect that the UV scale at which they set in is considerably
higher than the electroweak scale, leaving the naturalness problem unresolved.

There are a
number of experimental observations and theoretical questions, unrelated to the 
naturalness problem, 
that point to new high energy scales. Neutrino masses, gauge coupling 
unification, 
dark matter, and the expectation of a more 
fundamental theory of gravity are all expected to introduce new scales and as a 
consequence introduce an additive renormalization of the Higgs 
mass parameter. 
None of these arguments, however, \emph{necessarily} points to a 
new UV scale relevant to the hierarchy problem. Neutrinos could 
be Dirac fermions with tiny Yukawa couplings, gauge coupling unification
may not occur or may not imply new superheavy states,
and dark matter could well be related to 
the electroweak scale itself.

There is still the challenging question of quantum gravity.
We know that gravity does not make sense as a fundamental 
(perturbative) 
quantum field theory at short distances~\cite{Weinberg:1980kq}. 
Naively, the Planck scale is expected to correspond at least
roughly to a physical threshold where new massive states appear. 
In string theory this is indeed the case, and one can also argue that the
existence of microscopic black holes is enough to require fine-tuning
of the Higgs mass parameter \cite{Dvali:2012wq}.
A non-perturbative 
theory of quantum gravity might avoid this problem, but not if it resembles
strongly-coupled gauge theories where new massive states are connected 
to the scale of strong coupling. 
At present no mechanism is known
that can realize even a toy model for the type of UV completion that would 
avoid the hierachy problem, 
despite promising models in 2d~\cite{Dubovsky:2013ira}.

On the other hand, all claims about Planckian physics 
and resulting effects on the renormalization of the Higgs mass parameter are, 
at this point, speculative. Generic UV completions of the Standard Model 
certainly have a Higgs naturalness
problem \cite{Tavares:2013dga}, but
for all we know, spacetime geometry
breaks down at the Planck scale, and whether this results in a physical cut-off 
of relevance to the Higgs mass parameter is 
an open question. 
% Generic UV completions of the Standard Model certainly have a Higgs 
% naturalness
% problem \cite{Tavares:2013dga}, but the UV completion realized in nature
% is not generic, and at present we can only guess about its special features.

Following the same line of reasoning, it is not clear to what extent the 
existence of ultra-Planckian Landau poles, as occurs for the hypercharge
gauge coupling of the SM, should be 
regarded as a fundamental issue.
In particular, our semi-classical understanding of gravity seems to indicate 
that such Landau poles are unobservable; the requisite scattering experiments 
would 
presumably be dominated
by black hole production long before reaching the regime where incipient strong 
coupling in the hypercharge
interactions would show itself \footnote{This observation is from Steve 
Giddings.}.
It is striking that no couplings of the SM run into a Landau 
pole \emph{below} the Planck scale, which would be an unambigous sign of a new 
scale and therefore of the need (presumably) to fine-tune the Higgs mass.

In this paper, we assume that all explicit mass parameters vanish at the Planck 
scale, either as the consequence of the UV completed Planckian theory, or in 
spite of it. In addition we will focus on extensions of the SM that are weakly 
coupled and have
no vacuum instability below the Planck scale.

%%%%%%%%%%%%%%%%%%%%%%%%%%%%%%%%%%%%%%%%%%%%%%%%%%%%%%%%%%%%%%%%

%%%%%%%%%%%%%%%%%%%%%%%%%
\section{UV Stability and IR Instability from Dark Sectors}\label{sec:TeVDM}
%%%%%%%%%%%%%%%%%%%%%%%%%%

%%%%%%%%%%%%%%%%%%%%%%%%%%
\subsection{UV Stability}
%%%%%%%%%%%%%%%%%%%%%%%%%%

It is an intriguing observation about the Standard Model, that it seems to be 
consistent 
up to very high mass scales. Below the Planck scale, the only hint for New 
Physics within the SM itself is a possible instability of the electroweak 
vacuum triggered by the large top Yukawa.
In the SM, the observed Higgs mass of $m_h \simeq 125.5$~GeV 
implies a Higgs quartic coupling at the electroweak scale of around 
$\lambda^\text{SM}_H(m_t) \simeq 
0.254$~\cite{Buttazzo:2013uya,Degrassi:2012ry}. 
With this 
infrared boundary condition, assuming central values for $m_t$ and $\alpha_s$, 
the Higgs quartic coupling runs negative at scales 
around $10^{10}$~GeV and stays at a small negative value $\lambda_H \simeq 
-0.02$ up to the Planck scale, rendering the electroweak vacuum 
unstable~\cite{Buttazzo:2013uya,Degrassi:2012ry}\footnote{The electroweak 
vacuum in the SM is still meta-stable, i.e. its lifetime is larger than the age 
of the universe.}.

The instability can for example be overcome by the
extension of the SM by a complex scalar $\Sigma$ with portal coupling to the 
Higgs $H$, so 
that the most general scalar potential reads
\begin{align}\label{eq:SLag}
V(H,\Sigma)=\mu_H^2\,H^\dagger H + \frac{\lambda_{H}}{2}(H^\dagger 
H)^2 +
 \mu_\Sigma^2\Sigma^\dagger \Sigma+ \frac{\lambda_{\Sigma}}{2}(\Sigma^\dagger 
\Sigma)^2+
 \lambda_{\Sigma H}\Sigma^\dagger \Sigma H^\dagger H  \,.
\end{align}
The new scalar can affect the stability of the Higgs potential in two ways: (i) 
by changing the beta function of the Higgs quartic; (ii) by changing the 
infrared boundary condition of the Higgs quartic. We briefly review both 
possibilities.

(i) The portal coupling $\lambda_{\Sigma H}$ gives a positive contribution to 
the beta function of the 
Higgs 
quartic. At the one loop level we have
\begin{equation}
 \beta_{\lambda_H} =  \frac{1}{16\pi^2} \left\{ 12\lambda_H^2 -\lambda_H\left(3 
(g^\prime)^2 + 9 g^2\right) + \frac{3}{4}(g^\prime)^4 + \frac{3}{2} 
(g^\prime)^2g^2 + \frac{9}{4} g^4 +12\lambda_H Y_t^2 - 12Y_t^4 + 2 
\lambda_{\Sigma H}^2 \right\} ~,
\end{equation}
where $g^\prime$ and $g$ are the $U(1)$ and $SU(2)$ gauge couplings, $Y_t$ is 
the top Yukawa coupling and we neglected the contributions from all other 
Yukawa couplings. In the SM, the top Yukawa contribution, $- 12Y_t^4$, 
dominates 
at low scales and drives the Higgs quartic coupling negative.
If $\lambda_{\Sigma H}$ is sufficiently large, 
it can balance
the top contribution and stabilize the vacuum. 
If in addition the vacuum expectation value of the new scalar vanishes,
$\langle \Sigma \rangle =0$, the scalar can be stable and is a dark matter 
candidate 
\cite{Hambye:2007vf,Clark:2009dc,Lerner:2009xg,Gonderinger:2009jp}.
 
(ii) If the new scalar has a non-vanishing vev, 
$\langle \Sigma \rangle =w/\sqrt{2}$,
the tree level scalar mass matrix in the broken phase of  
\eqref{eq:SLag} reads 
\begin{align}\label{eq:mass1}
 \mathcal{M}^2=\begin{pmatrix}
              \lambda_H\,v^2& \lambda_{\Sigma H} \,w\, v\\[3pt]
              \lambda_{\Sigma H} \,w\, v & 
\lambda_\Sigma\,w^2
             \end{pmatrix}\,.
\end{align}
In the limit $\lambda_\Sigma w^2 \gg \lambda_H v^2$, the light Higgs-like mass 
eigenstate has a 
mass
\begin{align}\label{eq:lam}
m_h^2 =\left( \lambda_H-\frac{\lambda_{\Sigma H}^2}{\lambda_\Sigma} \right) v^2 
 +\mathcal{O}\left(\frac{v^4}{w^2}\right)\,.
\end{align}
In order to reproduce a Higgs mass of $125.5$~GeV, the value of $\lambda_H$ at 
the electroweak scale has to be larger than in the SM. In that way
the UV instability can be avoided~\cite{EliasMiro:2012ay,Lebedev:2012zw,
Batell:2012zw}.
Indeed, for a Higgs quartic that is about $7\%$ larger than in the SM, 
$\lambda_H(m_t)\simeq 0.273$,
the central value of the Higgs quartic remains 
positive up to high scales and vanishes around the Planck mass. 
Taking into account uncertainties in the running of $\lambda_H$ from the top 
mass and strong gauge coupling, a positive Higgs quartic at the $2\sigma$ level 
corresponds to 
\begin{equation}
0.259 \lesssim \lambda_H(m_t) \lesssim 0.288 ~.
\end{equation}
Interestingly 
enough, for such a range of boundary conditions not only the Higgs quartic, but 
also 
its beta function become zero at scales close to the Planck scale. 
The required size of the portal coupling to stabilize the potential in the UV is
considerably smaller than using mechanism (i). If the scalar quartic 
$\lambda_\Sigma$ is of the same order of the Higgs quartic $\lambda_H$, a 
portal coupling of $|\lambda_{\Sigma H}| \sim 0.05$ is sufficient.
Although the heavy scalar is unstable in scenario (ii), additional fields which 
get their 
mass from couplings to $\Sigma$ could explain dark matter.

As pointed out in 
\cite{EliasMiro:2012ay}, the correction in $\eqref{eq:lam}$ from the heavy 
scalar persists even in the decoupling limit, $w \to \infty$, so that both 
mechanisms to 
mitigate the vacuum instability do not point to a specific scale for the 
extra sector\footnote{For mechanism (i) to work, the scale has to be at least 
somewhat below $10^{10}$~GeV, the scale where the Higgs quartic crosses zero.}.
This situation is fundamentally different in models with classical scale 
invariance in the UV. If the Higgs mass parameter 
$\mu_H^2$ in \eqref{eq:SLag} is zero it will be  
generated by the vev of the extra scalar through the portal coupling
\begin{align}
 \lambda_{\Sigma H}\Sigma^\dagger \Sigma \,H^\dagger H \quad \to\quad  
 \frac{\lambda_{\Sigma H} w^2}{2}\,H^\dagger H\,.
\end{align}
In that case, the ratio of vacuum expectation values is controlled by the 
portal coupling (note that $\lambda_{\Sigma H}$ has to be negative to trigger a 
vev for the Higgs boson)
\begin{align} \label{eq:vevs}
 \frac{v^2}{w^2}= -\frac{\lambda_{\Sigma H}}{\lambda_H} ~,
\end{align}
and the correction in \eqref{eq:lam} {\it does} decouple for $w \to \infty$.
For a vanishing Higgs mass parameter $\mu_H^2=0$ and for the Higgs 
being the lightest scalar, the vev of the extra scalar is therefore
bounded from above. 
The requirement that the central 
value ($2\sigma$ upper bound) of $\lambda_H$ remains positive up to the Planck 
scale implies
\begin{equation}
 w \lesssim \left(\lambda_\Sigma\right)^{-\frac{1}{4}} \times 350 ~(470) ~ 
\text{GeV}~.
\end{equation}
Here, we worked in the limit $\lambda_\Sigma w^2 \gg \lambda_H v^2$ and neglect 
the tiny $\lambda_{\Sigma H}$ contributions to 
the running of $\lambda_H$.
In the limit in which $\lambda_{\Sigma} w^2 = \lambda_H v^2$ the bounds become 
$w \lesssim 3.5 (12.7)$~TeV.
This corresponds 
to the extreme case of maximal mixing between the Higgs and the dark scalar. As 
we will discuss in Section~\ref{sec:higgs}, the mixing
 is strongly constrained by collider bounds.

%%%%%%%%%%%%%%%%%%%%%%%%%%
\subsection{IR Instability}
%%%%%%%%%%%%%%%%%%%%%%%%%%

We now address the question of how to generate the vev for the scalar $\Sigma$.
In particular, if not only the Higgs mass parameter, 
but all scales in the potential \eqref{eq:SLag} vanish in the UV, 
$\mu_H^2=\mu_\Sigma^2=0$, the 
vev of the extra scalar can only be induced radiatively, either through strong 
dynamics, in which a new condensation scale induces a mass term
for the extra scalar~\cite{Holthausen:2013ota, Heikinheimo:2013fta, 
Kubo:2014ova}, or by a 
Coleman-Weinberg 
mechanism, in which the balance between the quartic and the one-loop 
corrections to the effective 
potential
determine the vev~\cite{Coleman:1973jx}. 
We will concentrate on the latter mechanism in the following.

In the limit of small Higgs portal coupling $|\lambda_{ \Sigma H}|\ll 
1$, we can consider the effective potential $V_\mathrm{eff}$ for the scalar 
independently from the Higgs boson.
Its general one loop form is given by
\begin{align} \label{eq:Veff}
V_\mathrm{eff}(s,\mu)=
%V_0(s,\mu)+V_1(s,\mu)=
\frac{1}{8}\lambda_\Sigma(\mu)\,
s^4+\frac{B(\mu)}{4}\,s^4\,\log\left(\frac{s^2}{\mu^2}\right)\,,
\end{align}
in which the scalar component $s$ in \eqref{eq:HSig} is treated 
as a background field and $\mu$ is the renormalization scale.
Subleading terms in $V_\mathrm{eff}$ that are proportional to 
the anomalous dimension of $s$ are suppressed.
This effective potential has a local minimum if
\begin{align}\label{eq:min1}
 \frac{\partial V_\mathrm{eff}(s,\mu)}{\partial s }\bigg\vert_{\mu=s=w}=0 
\,\qquad \Rightarrow \qquad B=- \lambda_\Sigma\,,
\end{align}
and 
\begin{align} \label{eq:cond2}
 \frac{\partial^2 V_\mathrm{eff}(s,\mu)}{\partial s^2 }\bigg\vert_{\mu=s=w}>0 
\,\qquad \Rightarrow \qquad B>0\,.
\end{align}
Since $B$ is a loop-suppressed function, it follows, that in the vicinity of 
the minimum, the quartic coupling needs to be small and negative for CW 
symmetry breaking to work. For the full potential, including terms 
proportional to the portal coupling, $\lambda_{\Sigma H}$, 
the condition $\lambda_\Sigma<0$ is replaced by~\cite{Sher:1988mj}
\begin{align}\label{eq:fullmin}
 4\lambda_H\,\lambda_\Sigma-\lambda_{\Sigma H}^2<0\,.
\end{align}
 As long 
as the portal coupling is small, $\lambda_{\Sigma H}^2 \ll 
\beta_{\lambda_\Sigma}^{(1)} \lambda_H$, this gives 
approximately the same constraint\footnote{This situation changes for models 
in which the two-loop 
contribution 
becomes relevant~\cite{Hill:2014mqa}.}. Further, the coefficient $B$ is 
related 
to the beta-function of $\lambda_\Sigma$ by the one-loop renormalization group 
equation
\begin{align}
 \mu\frac{\partial}{\partial \mu} V_1(s,\mu) 
+\beta^{(1)}_{\lambda_\Sigma}\frac{d}{d\lambda_\Sigma} 
V_0-\gamma_s^{(1)}s\frac{d}{ds}V_0=0~,
\end{align}
in which $\gamma_s^{(1)}$ denotes the one-loop anomalous dimension of the 
scalar fields $s$ and $\beta_{\lambda_\Sigma}^{(1)}$ the one-loop beta function 
of the 
quartic coupling.
From the general form~\eqref{eq:Veff} follows
\begin{align}
\beta_{\lambda_\Sigma}^{(1)}=4\gamma_s^{(1)}\,\lambda_\Sigma+ 4 B~.
\end{align}
Close to the minimum, the first term can be neglected to good approximation. 
Therefore the beta function of $\lambda_\Sigma$ has to be positive to induce a 
vev for the scalar.
A natural way to ensure a positive beta function in a region around the minimum 
is to charge the scalar under a dark gauge symmetry. Loops with dark gauge 
bosons give a positive contribution to $\beta_{\lambda_\Sigma}$ and lead to the 
desired IR instability of the scalar potential. At the same time, a positive 
beta function for the quartic coupling ensures that the scalar potential is 
stable in the UV.

%%%%%%%%%%%%%%%%%%%%%%%%%%
\subsection{The Scale of Dark Matter}
%%%%%%%%%%%%%%%%%%%%%%%%%%

Independent on whether the vev of the dark scalar is induced by strong dynamics 
or by a 
Coleman-Weinberg mechanism as discussed in the previous section, 
additional fields 
with couplings to $\Sigma$ are required, which can provide dark matter 
candidates. 
If the new sector 
does not introduce explicit mass scales, the masses of any new states can 
only be generated through the vev of the extra scalar. In this context it is 
very interesting that the range suggested by stability considerations seems to 
agree with the mass scales suggested by the ``WIMP miracle''. 
We emphasize that this is a generic feature of extensions of the SM with the 
above properties. Extended 
models with additional scalars can of course soften this relation.

Examples of models in the literature, in which the electroweak vacuum is 
stabilized through the Higgs portal and dimensionful couplings are absent, 
reveal the mentioned connection between the dynamical generation of a vacuum 
expectation value in the IR, the stabilization of the vacuum in the UV, and the 
dark matter sector:
The authors of~\cite{Iso:2012jn} discuss a model with 
an extra scalar charged under a $U(1)_\mathrm{B-L}$ that gives a Majorana mass 
to right-handed neutrinos through a CW mechanism. The $Z'$ in this model is not 
a candidate for DM, because it couples to $B-L$ and the resulting experimental 
bounds on the 
$Z'$ push the vev of the extra scalar above 3 TeV. As a consequence, the vacuum 
cannot be stabilized up to the Planck scale in this model. 
In \cite{Hambye:2013dgv}, the extra scalar is a doublet under an additional dark
$SU(2)$ and breaks it completely. In this case the heavy gauge boson triplet 
constitutes dark matter, and if vacuum stability is enforced, the vev of the 
scalar is bound to be at the TeV scale.
In addition, the authors of \cite{Khoze:2014xha} have shown, that the extra 
gauge couplings that drive the quartic of the extra scalar negative in the IR 
need to be 
of order one in these models in order to stabilize the vacuum, so that the 
masses of the corresponding gauge bosons are bound from below by $m = g\, w/2 
\gtrsim 
500$~GeV. 

If the hidden sector
in addition to scalars and gauge bosons has also fermionic degrees of freedom, 
they are 
generically required to be lighter than the gauge bosons. This can easily be 
understood from the 
fact that they enter the effective potential with a negative sign,
\begin{align}\label{eq:Veff2}
 V_\mathrm{eff}(s,\mu)=V_0(s,\mu)+\frac{1}{64\pi^2}\sum_{i=B,F} n_i 
m_i^4(s)\,\left[\log\frac{m_i^2(s)}{\mu^2}-C_i\right]\,,
\end{align}
where the sum goes over fermions (F) and bosons (B), $m_i(s)$ 
denotes the corresponding Higgs dependent masses, $n_i =\mp $ the number of 
fermionic/bosonic degrees of 
freedom, and the $C_i$ are renormalization scheme dependent constants. If the 
fermionic contributions dominate the one loop contributions to the effective 
potential \eqref{eq:Veff2}, the condition~(\ref{eq:cond2}) cannot be fulfilled. 
Hence, the effective potential is 
unbounded from below, i.e. the fermions generate a UV instability instead of a 
IR instability. Therefore, for the CW mechanism to work, the gauge bosons 
are generically heavier and fermions constitute dark matter. 
In the following 
sections we will discuss in detail an example of a model 
\begin{itemize}
 \item that does not contain any explicit mass scales,
 \item that utilizes the Coleman-Weinberg mechanism in a dark sector to induce 
spontaneous electroweak symmetry breaking through a Higgs portal,
 \item that is weakly coupled below the Planck scale, with all of the running 
scalar quartic couplings starting near zero at the Planck scale,
 \item that stabilizes the vacuum until the Planck scale, and 
 \item that contains fermionic dark matter with masses at or below the 
electroweak scale. 
\end{itemize}

%%%%%%%%%%%%%%%%%%%%%%%%%
\section{The Model} \label{sec:model}
We consider an extension of the SM by a $SU(2)_X \times U(1)_X$ gauge group, 
under which all SM fields are uncharged. In addition to the $SU(2)_X \times 
U(1)_X$ gauge bosons 
$W^\prime_a$ and $B^\prime$, we introduce a scalar doublet $\Sigma$ under 
$SU(2)_X$ with $U(1)_X$ charge $Q^X_\Sigma = 1/2$. The fermionic sector 
consists of two sets of chiral SM singlet fermions: $\psi_i^L$, 
$\xi_i^R$, $\chi_i^R$, with $i = 1,2$. 
The left handed fields $\psi_1^L = (\chi_1^L , \xi_1^L)$ and $\psi_2^L = 
(\xi_2^L , \chi_2^L)$ are $SU(2)_X$ 
doublets, while the right handed ones are $SU(2)_X$ singlets. We assign the 
following dark hypercharges that ensure anomaly cancellation: $Q^X_{\psi_1} = 
+1/2$, $Q^X_{\psi_2} = -1/2$, $Q^X_{\chi_1} = +1$, $Q^X_{\chi_2} = -1$, 
$Q^X_{\xi_1} = 
Q^X_{\xi_2} = 0$.\\
We denote the field strength 
tensors of the $SU(2)_X$ and $U(1)_X$ gauge symmetries by 
$(W^\prime_a)_{\mu\nu}$ and 
$(B^\prime)^{\mu\nu}$, so that their kinetic terms read
\begin{eqnarray} \label{eq:gaugelagrangian}
 \mathcal{L}_\text{gauge} =\frac{1}{4} (W^\prime_a)_{\mu\nu} 
(W^\prime_a)^{\mu\nu} + \frac{1}{4} (B^\prime)_{\mu\nu} (B^\prime)^{\mu\nu}~,
\end{eqnarray}
where $a = 1,2,3$ is the index of the adjoint of $SU(2)_X$.
We assume that there is no kinetic mixing between the $U(1)_X$ gauge boson and 
the SM hypercharge gauge boson.
As our model does not contain fields that are charged under both $U(1)$ 
symmetries, such a choice is stable under radiative corrections.
In the absence of kinetic $U(1)$ mixing, the only renormalizable portal between 
the dark sector and 
the SM is the mixing of the dark scalar with the Higgs.
Explicit mass terms for the scalars are assumed to vanish, such that
\begin{equation} \label{eq:scalarlagrangian}
 \mathcal{L}_\text{scalar} = |D\,H|^2 + |D\,\Sigma|^2 - \frac{\lambda_H}{2} 
|H|^4 
- 
\frac{\lambda_\Sigma}{2} |\Sigma|^4 - \lambda_{\Sigma H} |H|^2 |\Sigma|^2 ~.
\end{equation}
The covariant derivatives of $H$ and $\Sigma$ are given by 
(Lorentz indices are suppressed for simplicity)
\begin{equation}
 D\, H = (\partial - i \frac{g}{2} \sigma^a W_a - i g^\prime 
Q_H B) H ~,~~  D\,\Sigma = (\partial - i \frac{g_X}{2} \sigma^a W^\prime_a 
- 
i g_X^\prime 
Q^X_\Sigma B^\prime)\Sigma ~.
\end{equation}
Here,  $g$ and $g^\prime$ are the $SU(2)$ and $U(1)$ gauge couplings of the SM, 
and $g_X$ and $g_X^\prime$ are the corresponding couplings in the dark sector.
The Higgs and the dark scalar can be decomposed as follows
\begin{equation}\label{eq:HSig}
 H = \begin{pmatrix} G^+ \\ \frac{1}{\sqrt{2}}(h + v + iG^0) \end{pmatrix} 
~,~~~ \Sigma = \begin{pmatrix} a^+ \\ \frac{1}{\sqrt{2}}(s + w + 
ia)\end{pmatrix} ~,
\end{equation}
where $v$ ($w$) is the respective vacuum expectation value that breaks the 
$SU(2)_{(X)}\times U(1)_{(X)}$ gauge group down to (dark) electromagnetism. 
The Goldstone bosons $G^\pm$, $G^0$ and $a^\pm$, $a$ provide the longitudinal 
components of the $W$ and $Z$ boson of the SM, as well as the corresponding 
$W^\prime$ and $Z^\prime$ in the dark sector. 
The masses of the dark gauge bosons are given by
\begin{equation}\label{eq:Bmasses}
m_{\gamma'}=0~,~~ m_{W^\prime} = \frac{w}{2} g_X ~,~~ m_{Z^\prime} = 
\frac{w}{2} \sqrt{g_X^2 + 
g_X^{\prime~2}} ~.
\end{equation}
Analogous to the photon in the SM, the dark sector contains a massless 
gauge boson, which we will refer to as dark photon, $\gamma^\prime$.
In complete analogy to the SM, we define a dark electromagnetic coupling $e_X$ 
as well as a dark mixing angle $\theta_X$
\begin{equation}
 e_X = \frac{g_X g_X^\prime}{\sqrt{g_X^2 + g_X^{\prime~2}}} ~,~~ c_X = 
\cos\theta_X = \frac{g_X}{\sqrt{g_X^2 + g_X^{\prime~2}}} ~,~~ s_X = 
\sin\theta_X = \frac{g_X^\prime}{\sqrt{g_X^2 + g_X^{\prime~2}}} ~.
\end{equation}
The dark fermions couple to the extra scalar $\Sigma$ through Yukawa couplings, 
\begin{eqnarray} 
 \mathcal{L}_\text{fermion} &=& i \bar\psi_i^L D\!\!\!\!/ ~\psi_i^L + i 
\bar\chi^R_i D\!\!\!\!/ ~\chi^R_i + i \bar\xi^R_i \partial\!\!\!/ ~\xi^R_i 
\nonumber \\
&& + 
(Y_{\chi_1} \bar\psi^L_1 \chi^R_1 \tilde \Sigma + Y_{\chi_2} \bar\psi_2^L 
\chi_2^R 
\Sigma + Y_{\xi_1} \bar\psi^L_1 \xi^R_1 \Sigma + Y_{\xi_2} \bar\psi_2^L \xi_2^R 
\tilde \Sigma  ~+~\text{h.c.}  )~,\label{eq:fermionlagrangian}
\end{eqnarray}
where $\tilde \Sigma = i \sigma_2 \Sigma^*$.
As in the scalar sector, we do not consider explicit Majorana mass terms for 
the fermions
that would be allowed given the quantum number assignments for $\psi_i$, 
$\chi_i$, and $\xi_i$. 
For simplicity, we also choose flavor 
diagonal Yukawa couplings for the $\xi_i$ fields. Both the absence of Majorana 
masses and of flavor off-diagonal Yukawa couplings can for example be enforced 
by demanding dark fermion number conservation. The covariant 
derivatives of the fermions are 
\begin{equation}
 D\psi_i^L = (\partial - i \frac{g_X}{2} \sigma^a W^\prime_a - i g_X^\prime 
Q^X_{\psi_i} B^\prime)\psi_i^L ~,~~ D\chi_i^R = (\partial - i g_X^\prime 
Q^X_{\chi_i} B^\prime)\chi_i^R ~,
\end{equation}
and $\xi^R_i$ are total singlets. After breaking of the dark $SU(2)_X \times 
U(1)_X$ by the vev of $\Sigma$, the fermions become massive 
and we introduce 
the Dirac spinors $\chi_i = P_L \chi_i + P_R \chi_i = (\chi^L_i , \chi^R_i)$ 
and $\xi_i = P_L \xi_i + P_R \xi_i = (\xi^L_i , \xi^R_i)$ with masses
\begin{equation}\label{eq:fmasses}
 m_{\chi_i} = \frac{Y_{\chi_i}}{\sqrt{2}} w ~,~~ m_{\xi_i} = 
\frac{Y_{\xi_i}}{\sqrt{2}} w ~.
\end{equation}
Conservation of dark 
fermion number and dark electromagnetism implies that both $\chi_i$ and $\xi_i$ 
can be stable dark matter candidates.

%%%%%%%%%%%%%%%%%%%%%%%%%%%%%%%%%%%%%%%%%%%%%%%%%%%%%%%%%%
\subsection{The Scalar Spectrum}\label{sec:CW}
%%%%%%%%%%%%%%%%%%%%%%%%%%%%%%%%%%%%%%%%%%%%%%%%%%%%%%%%%%%

The one loop effective potential of the model is given in the 
Appendix~\ref{sec:Veff}.
If the bosonic contributions to the effective potential dominate over the 
fermionic ones, non-zero scalar vevs will be induced radiatively. In the limit 
of a small portal coupling, the vevs of the Higgs $v$ and of the dark scalar 
$w$ are approximately connected by the relation~(\ref{eq:vevs}). 
Neglecting the effects of the field anomalous dimensions of the Higgs and the 
dark scalar as well as the running of both the SM quartic coupling and the 
portal coupling, while keeping the dominant contribution from the running of 
$\lambda_\Sigma$, the scalar mass matrix in the minimum of the potential can be 
written as
\begin{equation}\label{eq:mass2}
 \mathcal{M}^2 \simeq \frac{v^2}{2} \begin{pmatrix} 2\lambda_H & -2 
\sqrt{\lambda_H |\lambda_{\Sigma H}|}\\ -2 \sqrt{\lambda_H |\lambda_{\Sigma 
H}|} & 
2 |\lambda_{\Sigma 
H}| + \lambda_H \beta_{\lambda_\Sigma}/ |\lambda_{\Sigma H}|\end{pmatrix} ~.
\end{equation} 
This mass matrix can be diagonalized through the rotation
\begin{equation}\label{eq:sina}
 \begin{pmatrix} h \\ s \end{pmatrix} \rightarrow \begin{pmatrix} c_\alpha & 
s_\alpha \\ -s_\alpha  & c_\alpha \end{pmatrix} \begin{pmatrix} h \\ s 
\end{pmatrix} ~,~~ \sin2\alpha = \frac{2 \sqrt{\lambda_H|\lambda_{\Sigma H}|} 
v^2}{m_s^2 - m_h^2} ~,
\end{equation}
with $s_\alpha = \sin\alpha$ and $c_\alpha = \cos\alpha$.
The mass eigenvalues $m_h$ and $m_s$ are given by
\begin{equation} \label{eq:masses}
 m_h^2 \simeq v^2 \left( \lambda_H - \frac{2 \lambda_{\Sigma 
H}^2}{\beta_{\lambda_\Sigma} - 2 |\lambda_{\Sigma H}|} \right) ~,~~ m_s^2 
\simeq 
v^2 \left( \frac{\lambda_H \beta_{\lambda_\Sigma}}{2|\lambda_{\Sigma H}|} + 
\frac{\beta_{\lambda_\Sigma}|\lambda_{\Sigma H}|}{\beta_{\lambda_\Sigma} - 2 
|\lambda_{\Sigma H}|} 
\right) ~,
\end{equation}
where we expanded to first order in the limit $\lambda_{\Sigma H} , 
\beta_{\lambda_\Sigma} \ll \lambda_H$. 
In this limit, the mass of the dark scalar is directly proportional to the beta 
function of the dark scalar quartic coupling. If the dark scalar beta function 
is larger than twice the absolute value of the portal coupling, 
\begin{equation} \label{eq:scalarmass}
 \beta_{\lambda_\Sigma} \gtrsim 2 |\lambda_{\Sigma H}| ~,
\end{equation}
the dark scalar is heavier than the Higgs boson, and the mass of the Higgs 
boson is reduced compared to the Standard Model expression.

%%%%%%%%%%%%%%%%%%%%%%%%%%%%%%%%%%%%%%%%%%%%%%%%%%%%%%%%%%%
\subsection{Vacuum Stability in the UV} \label{sec:RGE}
%%%%%%%%%%%%%%%%%%%%%%%%%%%%%%%%%%%%%%%%%%%%%%%%%%%%%%%%%%%

We now discuss the renormalization group running of the model parameters up to 
high scales and demonstrate that the electroweak minimum in the scalar 
potential can be absolutely stable. 
The one loop beta functions for all couplings of the dark sector as well as the 
one loop correction to the beta function of the Higgs quartic are collected in 
the Appendix~\ref{sec:betafunctions}.
For the SM beta functions we use 2 loop results from~\cite{Machacek:1983tz,
Machacek:1983fi,Machacek:1984zw,Luo:2002ey,Luo:2002ti,Luo:2002iq}.

%%%%%%%%%%%%%%%%%%%%%%%%%
\begin{figure}[tb]
\centering
\includegraphics[width=0.6\columnwidth]{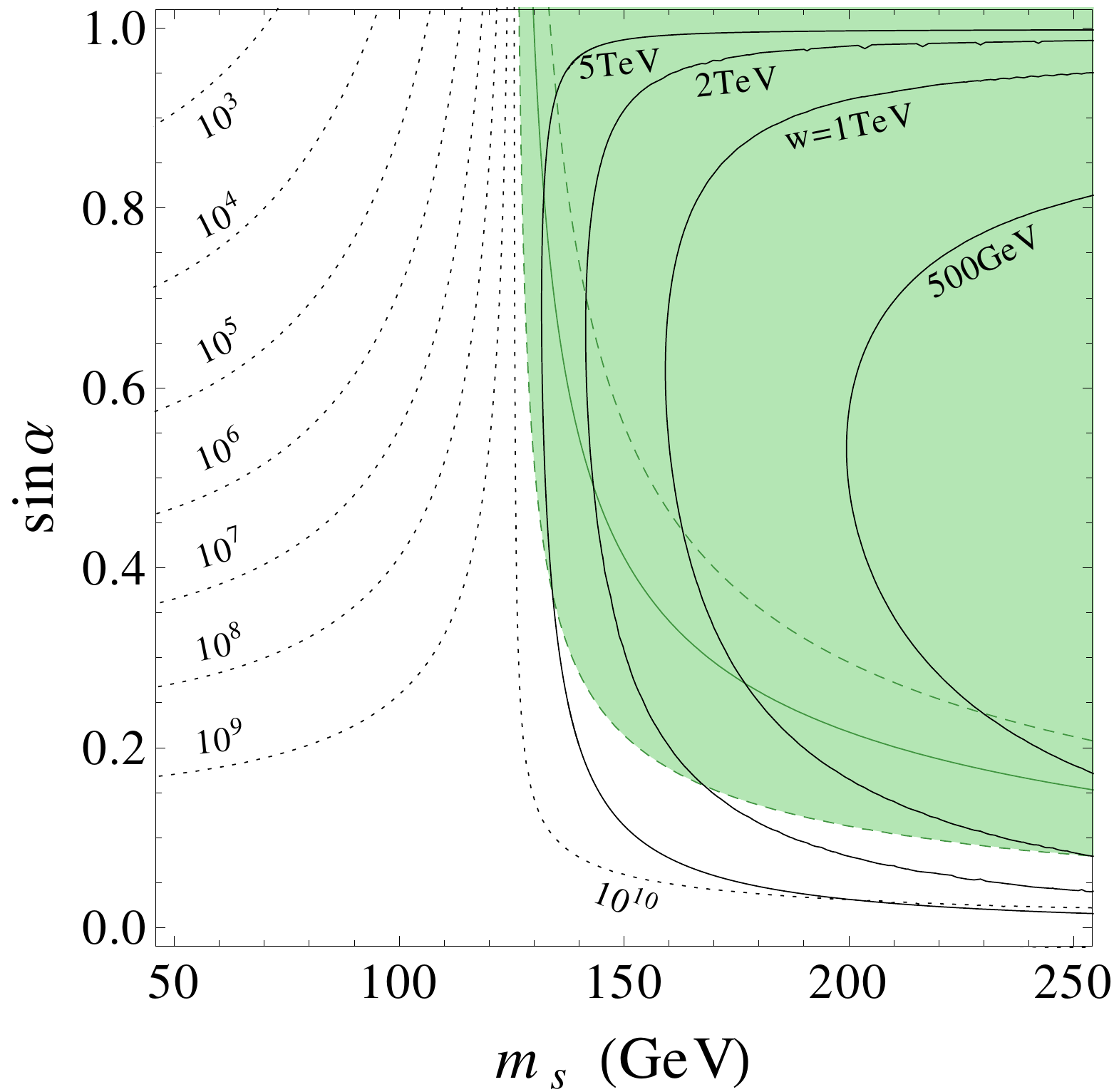}
\caption{Vacuum stability properties in the $m_s$-$\sin\alpha$ plane. In the 
shaded region the Higgs quartic is positive up to the Planck scale. Between the 
two dashed contours the Higgs quartic touches zero close to the Planck scale 
within $2\sigma$. The dotted lines in the unstable region show the scale at 
which the Higgs quartic runs negative. The solid lines indicate contours of 
constant scalar vev, $w$. Note, that large mixing angles $\sin\alpha \gtrsim 
0.5$ are phenomenologically strongly constrained by collider bounds, see 
Section \ref{sec:higgs}. 
\label{fig:stability}}
\end{figure}
%%%%%%%%%%%%%%%%%%%%%%%%%

As discussed already in Section~\ref{sec:TeVDM} and as shown in 
Equation~(\ref{eq:masses}), the physical Higgs mass is not 
completely determined by the Higgs quartic, but gets an additional contribution 
from the mixing with the dark scalar. If the dark scalar is heavier than the 
Higgs, 
mixing effects will reduce the Higgs mass and a quartic coupling larger than in 
the SM is required to accommodate a Higgs mass of $m_h \simeq 125.5$~GeV. If 
the portal coupling is large enough, the IR boundary condition for $\lambda_H$ 
is such that $\lambda_H$ stays positive all the way to the Planck scale, or, in 
the limiting case, ``touches'' zero close to the Planck scale. 
The region of the parameter space where this can be achieved is shown in 
Figure~\ref{fig:stability} in the plane of the scalar mass $m_s$ and the mixing 
angle $\sin\alpha$. In the shaded region the Higgs quartic is positive up to 
the Planck scale. Between the two dashed curves the limiting case where 
$\lambda_H$ touches zero close to the Planck scale can be realized within 
$2\sigma$.
The dotted lines in the unstable region indicate the scale in GeV where the 
Higgs quartic runs negative.
The solid lines show contours of constant scalar vev $w$, that corresponds to a 
given scalar mass $m_s$ and mixing angle $\sin\alpha$. 

As we will discuss in Section~\ref{sec:higgs}, the mixing angle is bounded at 
the order of $\sin\alpha \lesssim 0.5$. This implies a typical value for $w$ 
around the TeV scale, 
and an upper bound of several TeV, as expected from the general discussion in 
Section~\ref{sec:TeVDM}. On the other hand,
scalar vevs considerably below a TeV can in principle be achieved by increasing 
$|\lambda_{\Sigma H}|$ (see Equation~(\ref{eq:vevs})). However, this requires 
that the 
beta function $\beta_{\lambda_\Sigma}$ needs to to be increased simultaneously 
due to the 
bound~(\ref{eq:scalarmass}). The beta function $\beta_{\lambda_\Sigma}$ is 
also bounded from above by perturbativity requirements on the dark gauge 
couplings. As a result, values for $w$ considerably below the TeV scale are
disfavored.  
In the following, we will concentrate on regions of parameter space with 
$w=\mathcal{O}$(1~TeV) 
and a Higgs quartic that touches zero close to the Planck scale.

As long as the dark fermion Yukawa couplings are not too large, the beta 
function of the scalar quartic $\beta_{\lambda_\Sigma}$ is dominated by the 
dark gauge couplings and stays positive. In such a case, $\lambda_\Sigma$ 
increases monotonically with the RG scale and is always positive in the UV. 
Note, however, that sizable dark fermion Yukawas can modify the behavior of 
$\lambda_\Sigma$ in the UV. In particular, the model allows to accommodate the 
limiting case where not only the Higgs 
quartic but also the dark scalar quartic touches zero close to the Planck 
scale. In 
the approximation 
$\beta_{\lambda_\Sigma}\approx 4B$, it is straight forward to compute the 
leading terms in the beta function of the scalar quartic from 
comparing \eqref{eq:Veff2} with \eqref{eq:Veff}, using \eqref{eq:Bmasses} and 
\eqref{eq:fmasses},
\begin{align}
 \beta_{\lambda_\Sigma}^{(1)}\approx  \frac{1}{16\pi^2} 
\left\{ \frac{9}{4} g_X^4 + \frac{3}{2} g_X^2 g_X^{\prime~2} + \frac{3}{4} 
g_X^{\prime~4} - 4(Y_{\chi_1}^4 + 
Y_{\chi_2}^4 + Y_{\xi_1}^4 + Y_{\xi_2}^4) \right\} ~.
\end{align}
The full one loop expression for the beta function can be found in 
Appendix~\ref{sec:betafunctions}.
The beta function of $\lambda_\Sigma$ receives contributions dominantly from 
three sources: (i) from $SU(2)_X$ gauge boson loops, (ii) from fermion loops, 
and (iii) from $U(1)_X$ gauge boson loops. The gauge boson (fermion) loops 
increase (decrease) 
$\lambda_\Sigma$ for higher scales. 
At low scales, the $SU(2)_X$ contribution dominates and leads to the infrared 
instability in the scalar potential as discussed above. 
With the given particle content, the $SU(2)_X$ gauge interactions are 
asymptotically free. Therefore the contribution of the $SU(2)_X$ gauge bosons 
to the running of the scalar quartic becomes smaller and smaller for higher 
scales. At sufficiently high 
scales, the dominant contributions to the beta function can come from the 
fermion Yukawa couplings, and the scalar quartic will start to decrease again. 
Finally, at scales close to the Planck scale, the $U(1)_X$ gauge coupling, 
having a positive beta function, can 
become large and compensate the effect of the Yukawa couplings. It is possible 
to adjust parameters such that the scalar quartic as well as its beta function 
vanish exactly at the Planck scale.

%%%%%%%%%%%%%%%%%%%%%%%
%
\begin{figure}[tb]
\centering
\includegraphics[width=0.46\columnwidth]{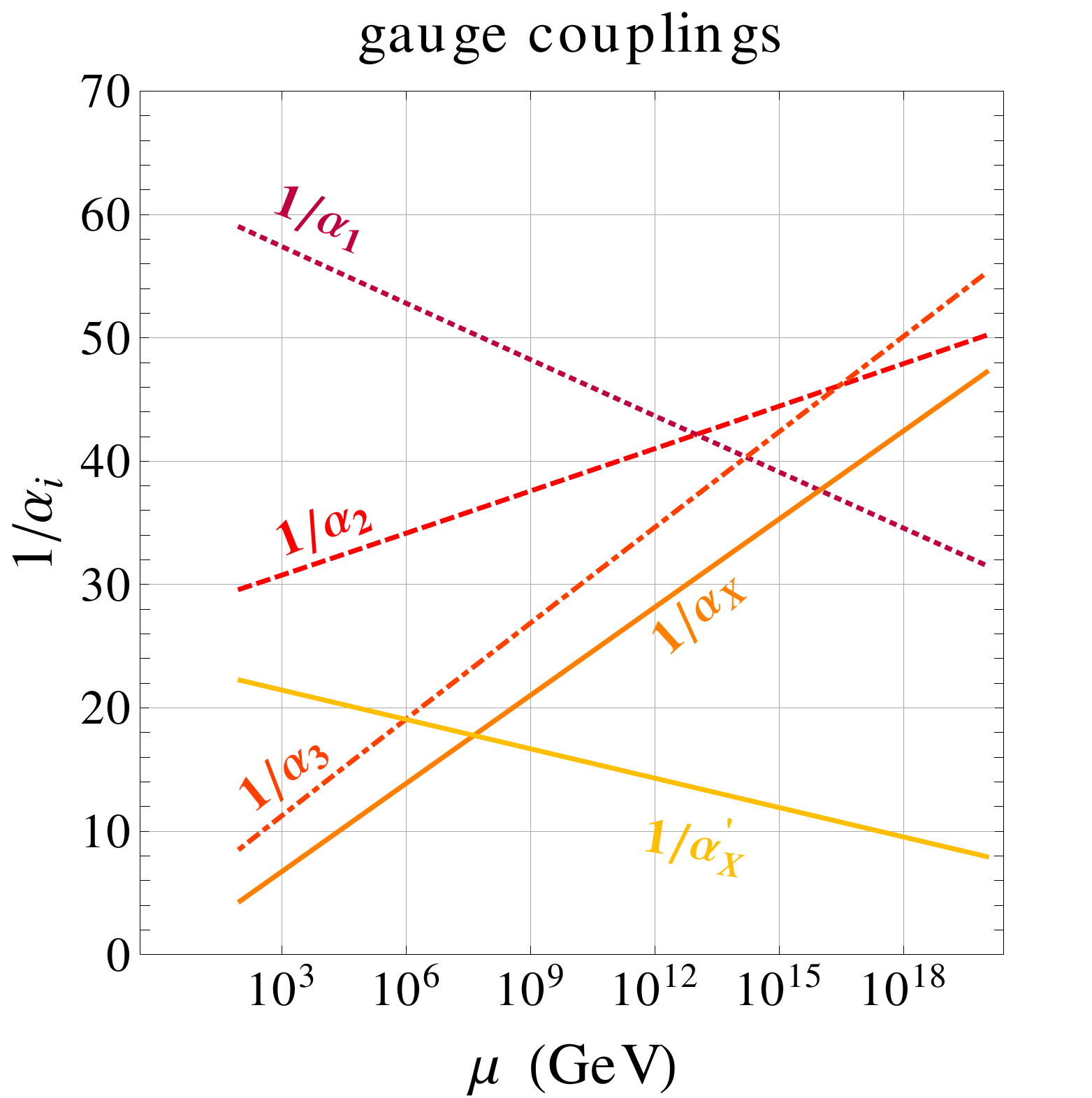} ~~~~~
\includegraphics[width=0.46\columnwidth]{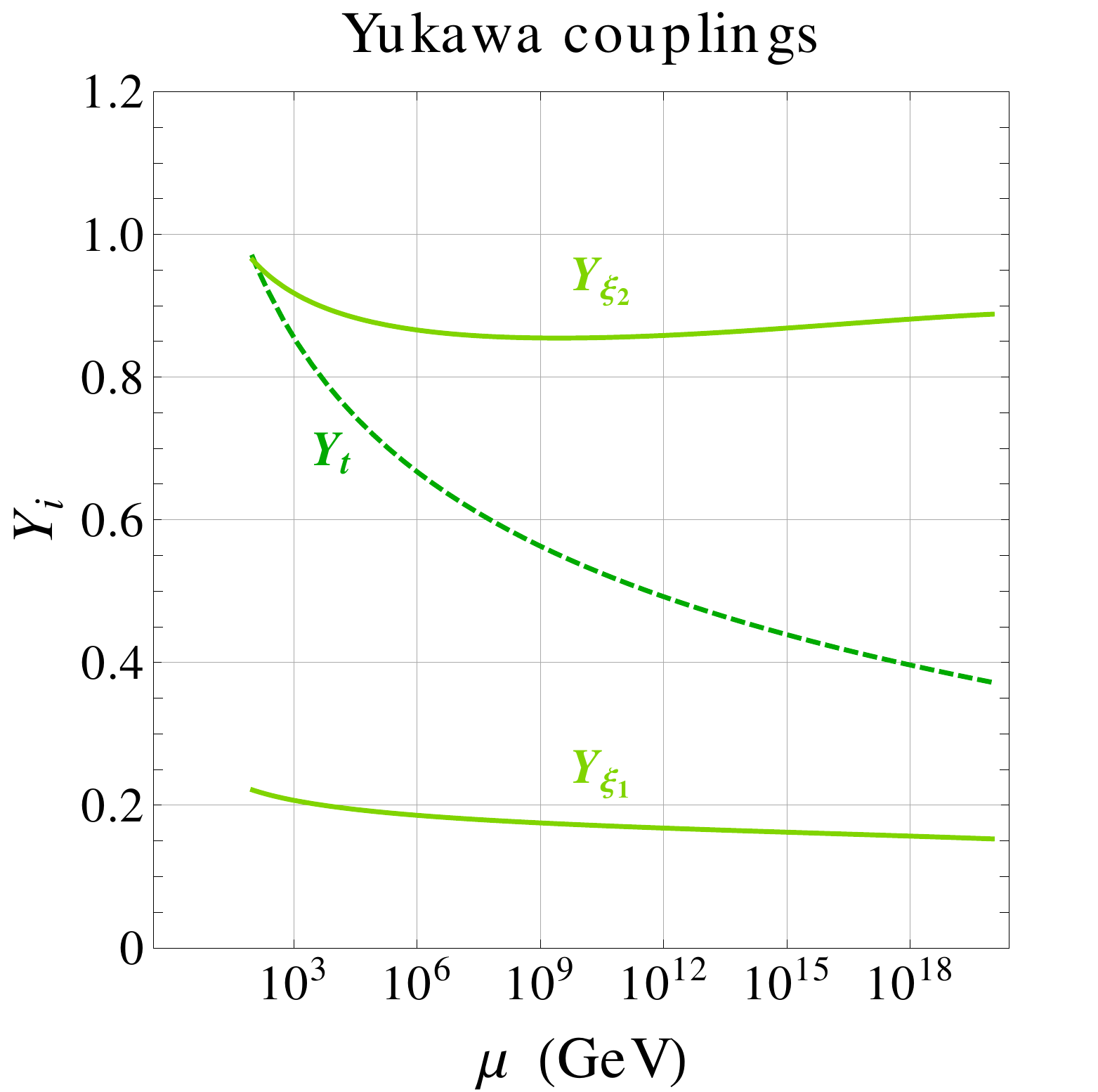} \\[20pt]
\includegraphics[width=0.46\columnwidth]{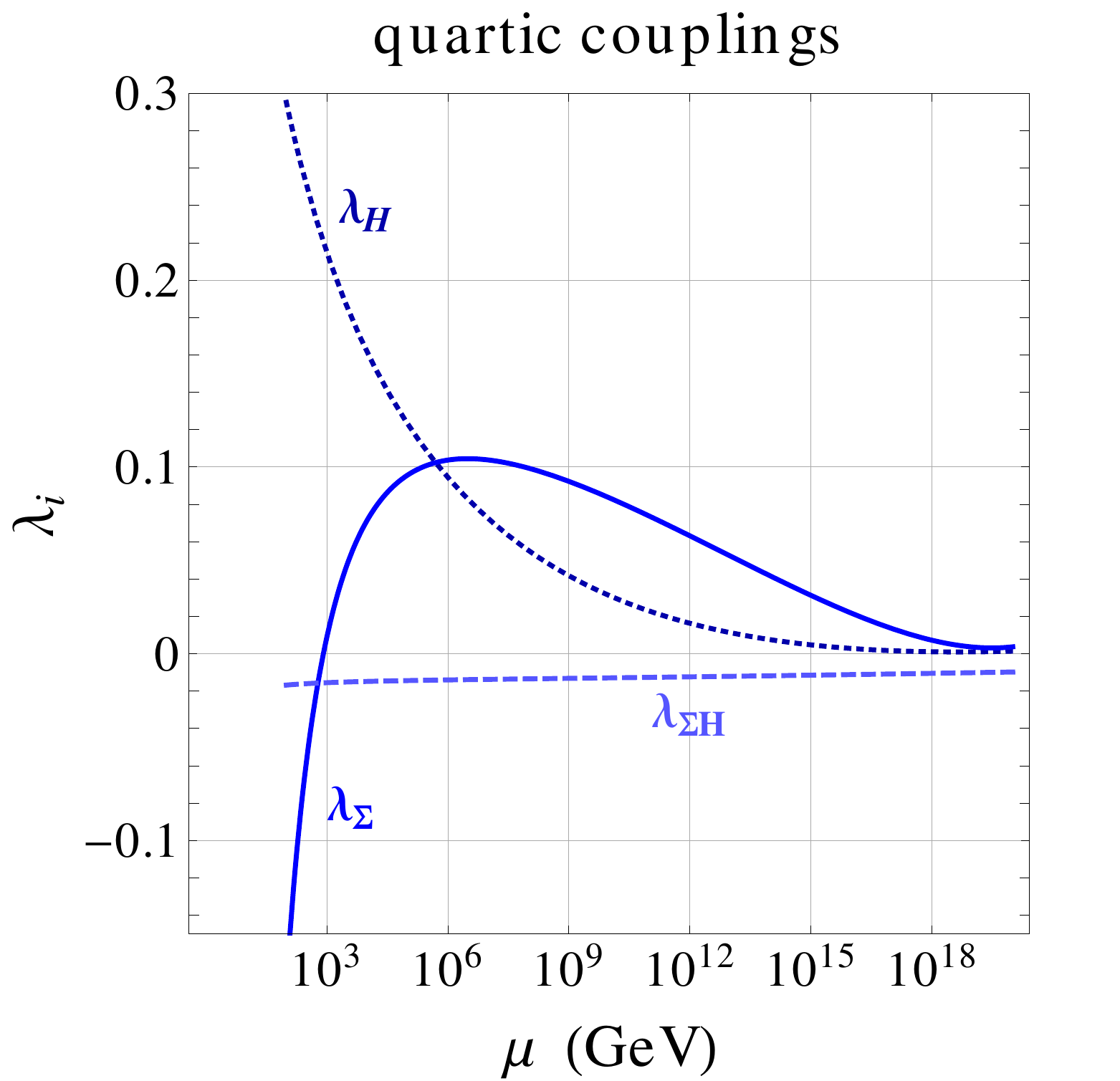}
\caption{The renormalization group evolution of the gauge couplings, the Yukawa 
couplings and the scalar quartic couplings for one example parameter point in 
the 
considered model that leads to an almost flat scalar potential at the Planck 
scale.}
\label{fig:RGE}
\end{figure}
%
%%%%%%%%%%%%%%%%%%%

The plots of Figure~\ref{fig:RGE} show the renormalization group evolution of 
the 
gauge couplings, Yukawa couplings and the scalar quartic couplings for an 
example parameter point of the model where such a limiting case is 
realized\footnote{Note that due to the small negative portal coupling 
$\lambda_{\Sigma 
H}$, the vacuum is actually not absolutely stable in the shown example. 
Absolute stability requires that $\lambda_H$ and $\lambda_\Sigma$ are at least 
of the same size as the (tiny) absolute value of the portal coupling.}.
The shown couplings correspond approximately 
to the following dark sector spectrum
\begin{eqnarray}
 m_h &\simeq& 125.5~ \text{GeV} ~,~~ m_s \simeq 168~ \text{GeV} ~,~~ 
m_{W^\prime} \simeq 740~
\text{GeV} ~,~~ m_{Z^\prime} \simeq 850~ \text{GeV} ~, \nonumber \\
 m_{\chi_1} &\simeq& 50~ \text{GeV} ~,~~~~ m_{\chi_2} \simeq 50~ \text{GeV} 
~,~~~~~ m_{\xi_1} 
\simeq 160~ \text{GeV} ~,~~ m_{\xi_2} \simeq 700~ \text{GeV} ~. 
\label{eq:darkmasses}
\end{eqnarray}
The values for the dark vev is
\begin{equation}
 w \simeq 1.1~ \text{TeV}~, 
\end{equation}
and the masses in~(\ref{eq:darkmasses}) (apart from the Higgs mass) correspond 
to running 
$\overline{\text{MS}}$ masses at the scale $\mu = w$.
The $SU(2)_X$ gauge coupling is $\mathcal{O}(1)$ at the low scale, while the 
$U(1)_X$ 
coupling is $\mathcal{O}(1)$ close to the Planck scale. The $U(1)_X$ gauge 
coupling
develops a Landau pole at around $10^{30}$~GeV, well above the Planck scale. 
Both the Higgs quartic $\lambda_H$ and the dark scalar quartic $\lambda_\Sigma$ 
as well as their beta functions are approximately 0 at the Planck scale. The 
portal coupling $\lambda_{\Sigma H}$ is small and negative at all scales but 
cannot run to zero at the Planck scale. It is the only link between the SM and 
the dark sector and is therefore only multiplicatively 
renormalized.

%%%%%%%%%%%%%%%%%%%%%%%%%
\section{Higgs and Dark Scalar Phenomenology \label{sec:higgs}}
%%%%%%%%%%%%%%%%%%%%%%%%%

The considered model leads to various testable predictions for Higgs 
phenomenology.
Due to the mixing of the two scalars, the couplings of the Higgs boson $h$ to 
all SM particles are suppressed by a factor $c_\alpha$ compared to the SM 
case, resulting in an overall suppression of all Higgs rates by $c_\alpha^2$. 
The latest results from Higgs rate measurements from 
ATLAS~\cite{ATLAS-CONF-2014-009} and CMS~\cite{CMS-PAS-HIG-14-009} read
\begin{equation}
 \mu_\text{ATLAS} = 1.30^{+0.18}_{-0.17} ~,~~ \mu_\text{CMS} = 1.00 \pm 
0.09^{+0.08}_{-0.07} \pm 0.07 ~, 
\end{equation}
which we will interpret roughly as a constraint of $c_\alpha \gtrsim 0.9$, 
equivalent to a $20\%$ reduction of the SM production rate.
At the next run of the LHC, the precision of the rate measurements is expected 
to be improved by around a factor of 3, which will allow to probe deviations of 
$c_\alpha$ from unity of the order of $5\%$.

Moreover, the mixing of the Higgs with the dark scalar also leads to couplings 
of $h$ to the fermions in 
the dark sector. If some of these fermions are sufficiently light, the Higgs 
can decay into them. We find for the corresponding partial decay widths 
\begin{equation} \label{eq:hff1}
 \Gamma(h\to f_i f_i) = \frac{Y_{f_i}^2}{8\pi} ~m_h~ s_\alpha^2 \left( 
1 - \frac{4 m_{f_i}^2}{m_h^2} \right)^{\frac 3 2}~,
\end{equation}
which applies for dark-charged and neutral fermions $f_i= \chi_i, \xi_i$.
Analogous to the SM decay of the Higgs into two photons, the Higgs can also 
decay into two dark photons through loops of dark-charged fermions $\chi_i$ and 
the dark $W^\prime$ boson. In the limit $m_{W^\prime} \gg m_h$, we find for the 
$h \to \gamma^\prime \gamma^\prime$ decay width
\begin{equation}\label{eq:hgaga}
 \Gamma(h\to \gamma^\prime \gamma^\prime) \simeq \frac{1}{16\pi} \, s_\alpha^2 
\frac{m_h^3}{w^2} \left(\frac{g_X^2}{16\pi^2}\right)^2 \left| 7 - \sum_i 
\frac{8 m_{\chi_i}^2}{m_h^2} \left[ 1 + \left( 1 - \frac{4 
m_{\chi_i}^2}{m_h^2}\right) f\left(\frac{m_h^2}{4 m_{\chi_i}^2}\right) \right] 
\right|^2 ~.
\end{equation}
The loop function $f$ is given in the Appendix~\ref{sec:loop}. 
Given that $h \to \gamma^\prime \gamma^\prime$ is loop suppressed, it can only 
compete with the decay into dark fermions if the dark gauge coupling is large 
$g_X \gtrsim 1$ and the fermion Yukawas are very small, $Y_{\chi_i} \lesssim 
10^{-2}$.

%%%%%%%%%%%%%%%%%%%%%%%
%
\begin{figure}[tb]
\centering
\includegraphics[width=0.46\columnwidth]{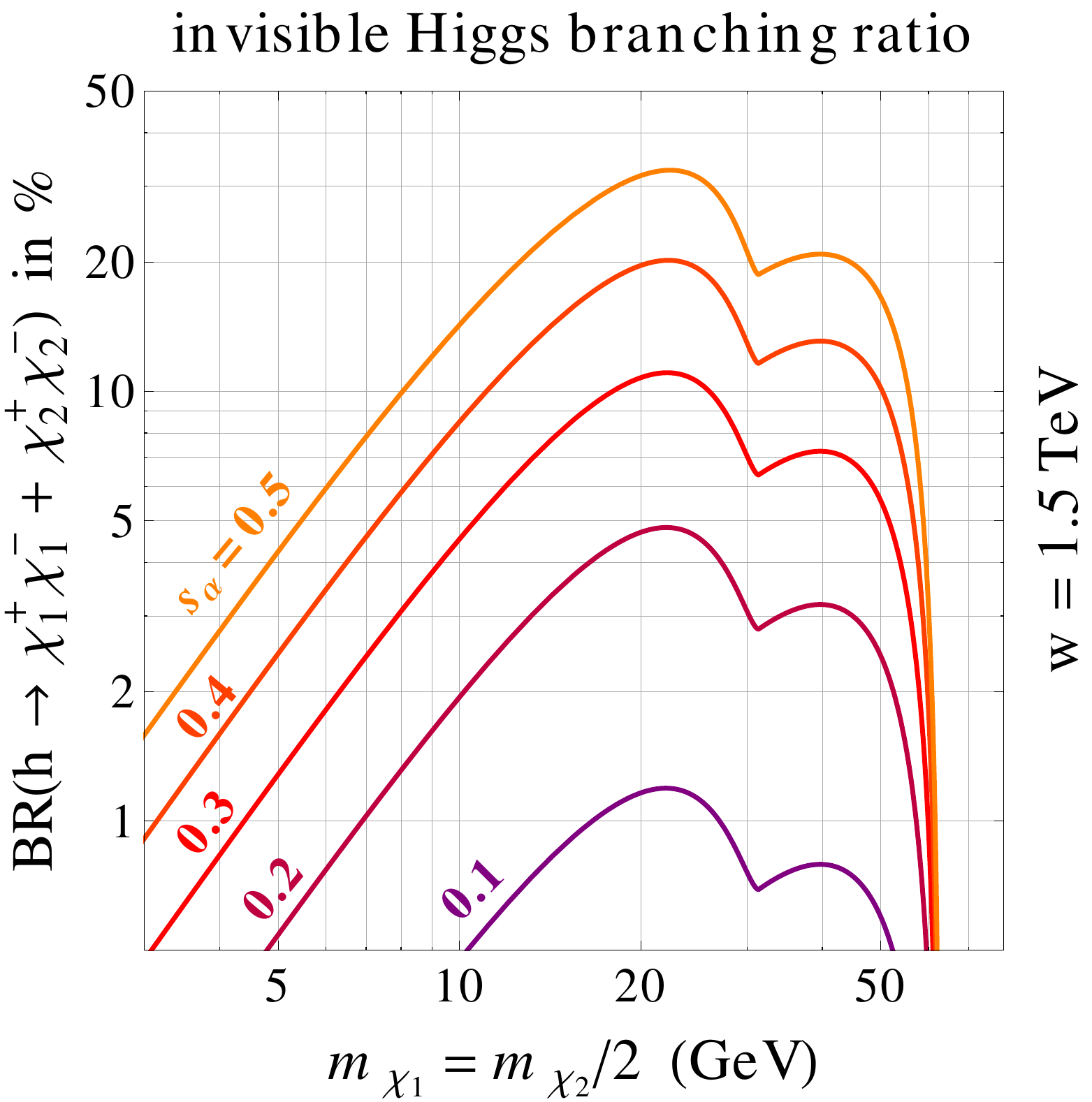} \\[20pt]
\includegraphics[width=0.46\columnwidth]{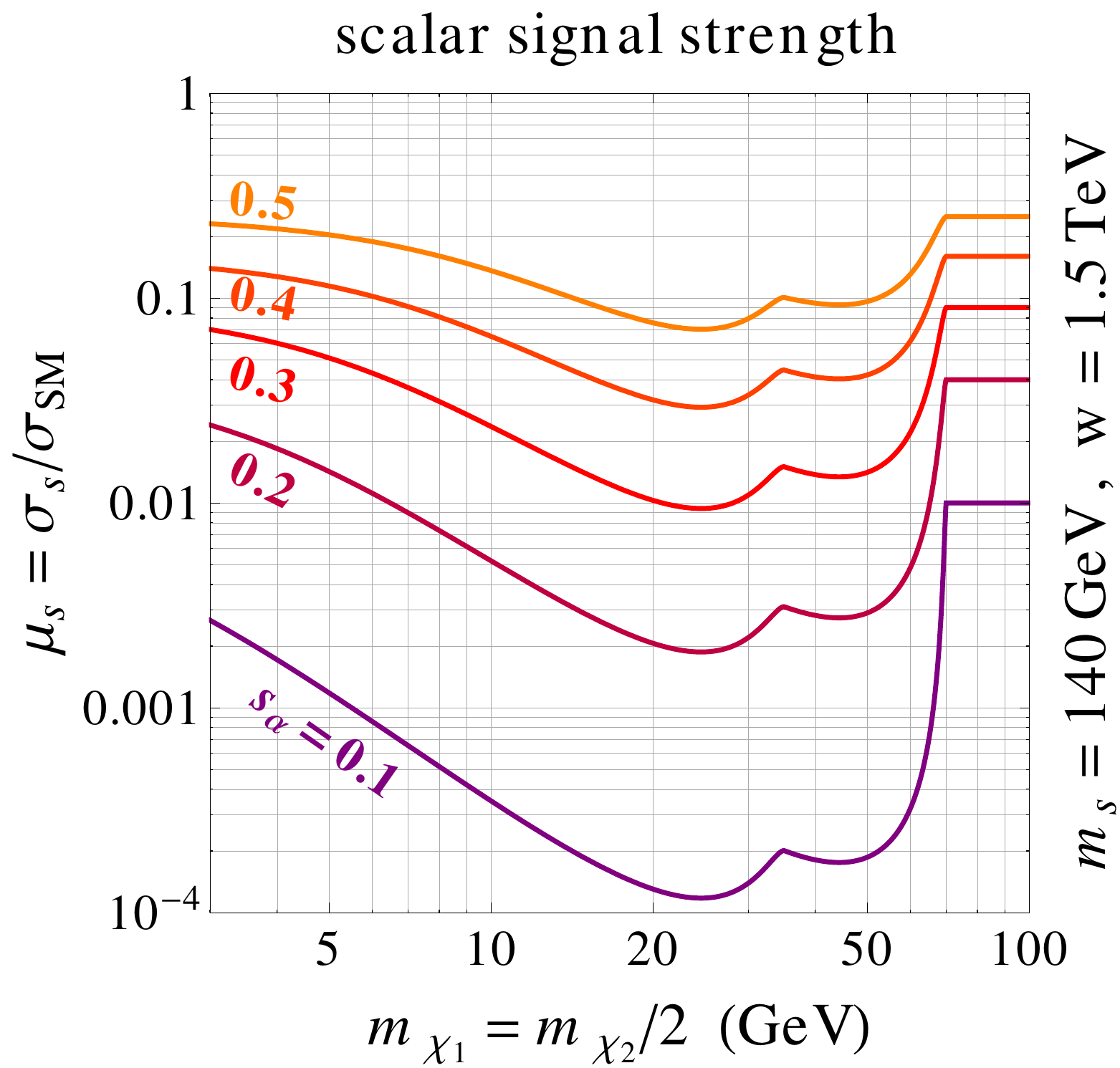} ~~~~~
\includegraphics[width=0.46\columnwidth]{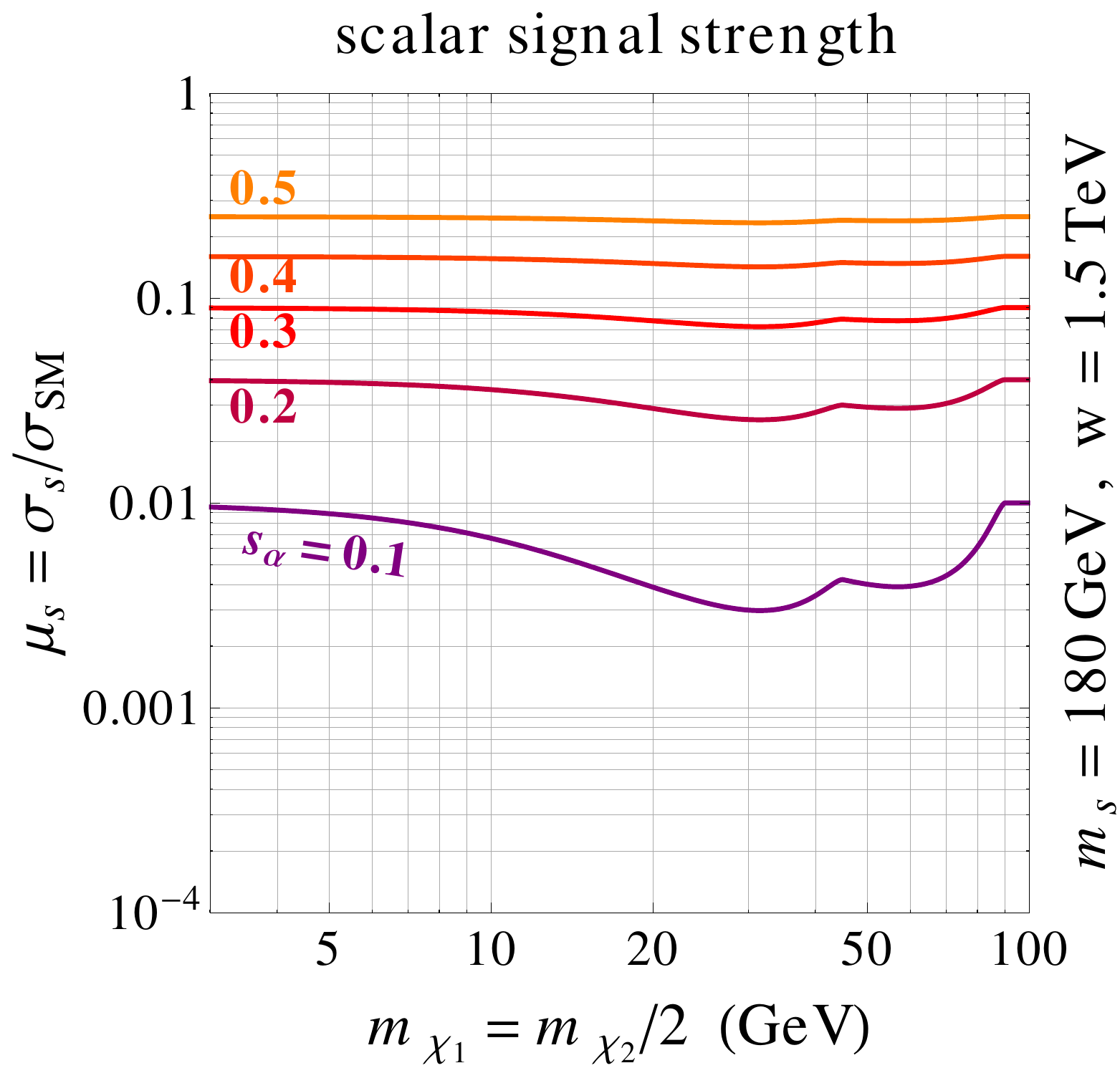}
\caption{Top: the invisible branching ratio of the Higgs boson as a function 
of the charged dark fermion 
mass, for example choices of the scalar mixing angle. Bottom: the 
scalar signal strength into SM particles as function of the charged dark 
fermion mass for example 
choices of the scalar mixing angle. The scalar mass is fixed to $m_s = 140$~GeV 
in the left and $m_s = 180$~GeV in the right plot.}
\label{fig:decay}
\end{figure}
%
%%%%%%%%%%%%%%%%%%%

Given the tiny total width of the SM Higgs, $\Gamma_h^\text{SM} 
\simeq 4$~MeV, even for moderate mixing angles $s_\alpha$ the induced invisible 
branching ratio can 
be sizable. This is illustrated in the upper plot of Figure~\ref{fig:decay}, 
that 
shows for various mixing angles $s_\alpha$ the branching ratio of the Higgs 
into the charged dark fermions as a function of the charged 
dark fermion mass, for the example choice $m_{\chi_1} = m_{\chi_2}/2$. The dark 
vev is set to $w = 1.5$~TeV and the neutral fermions are assumed to be heavier 
than at least half the Higgs mass. We observe that for moderate mixing angles 
of $s_\alpha \sim 0.3$, branching ratios into dark fermions of 
$\mathcal{O}$(10\%) are 
possible. The branching ratio can be even larger for smaller $w$.
The branching ratio into dark photons is at most at the percent level and 
therefore hardly relevant.  
ATLAS and CMS search for invisible decays of Higgs bosons that are produced in 
association with a $Z$ boson~\cite{Aad:2014iia,Chatrchyan:2014tja} and in 
vector boson fusion~\cite{CMS-PAS-HIG-13-013}.
The current best bound reads BR$(h \to \text{invisible}) \lesssim 58 
\% ~@95\%$~C.L.~\cite{Chatrchyan:2014tja} (see also~\cite{Zhou:2014dba} where a 
slightly stronger bound BR$(h \to \text{invisible}) \lesssim 40
\% ~@95\%$~C.L. has been obtained, recasting a CMS stop 
search~\cite{Chatrchyan:2013xna}). Bounds are expected to be improved 
down to BR$(h \to 
\text{invisible}) \lesssim 10 \%$ at the high luminosity 
LHC~\cite{Dawson:2013bba}.

Due to the mixing with the Higgs boson, the dark scalar $s$ acquires in turn 
couplings 
to all SM particles that are suppressed by a factor $s_\alpha$ compared to 
the 
SM Higgs. Therefore, the dark scalar can be searched for at the LHC in the 
usual Higgs searches. Particularly strong constraints arise already from 
current searches in 
the $WW$ and $ZZ$ channels that exclude a signal strength of the order of $\mu 
\sim 0.1$ over a very broad range of masses~\cite{Chatrchyan:2013iaa, 
Chatrchyan:2013mxa, ATLASC1, ATLASC2}. The production cross 
section of the scalar is suppressed by $s_\alpha^2$ with respect to a SM Higgs 
boson with the same mass. Therefore, we generically expect a bound on the 
mixing 
angle of the order of $s_\alpha \lesssim 0.3$. This is slightly more stringent 
than 
the bound obtained from Higgs rate measurements, $c_\alpha \gtrsim 0.9$, 
discussed above. 

Note, however, that also the dark
scalar can decay into dark sector particles. 
The corresponding partial width into dark fermions and dark photons are given 
by the expressions in~(\ref{eq:hff1}) and~(\ref{eq:hgaga}) 
with the replacements $m_h \to m_s$ and $s_\alpha \to c_\alpha$. 
If the scalar is light, with a mass 
below 
the $WW$ threshold, its decay width into SM particles is very small. Therefore, 
its invisible 
branching ratio can be sizable, in particular if the decay into dark fermions 
is kinematically accessible. This can easily reduce the branching ratio into 
SM 
particles by a factor of few or more and reduce the scalar signal strength well 
below $\mu_s 
= 0.1$ also for mixing angles of $s_\alpha \gtrsim 0.3$. This is illustrated in 
the 
lower left plot of Figure~\ref{fig:decay} that shows the signal strength of a 
$140$~GeV dark scalar for several choices of $s_\alpha$ as a function of the 
charged dark fermion masses $m_{\chi_1} = m_{\chi_2}/2$. If the dark scalar has 
a mass above the $WW$ threshold, its width is dominated by decays into $WW$ and 
decays into dark fermions tend to give only a small correction. This is 
illustrated in the lower right plot of Figure~\ref{fig:decay}, where we show 
the signal 
strength of the dark scalar for a dark scalar mass of $m_s = 180$~GeV.

In summary, we find that in the bulk of parameter space the prospects for 
detecting the dark scalar at the next run of the LHC are excellent, unless in 
the case where it dominantly decays into dark fermions. In the latter case, 
precision measurements of the Higgs signal strength in inclusive Higgs 
production will provide the strongest 
constraint on the mixing angle. In the case of sufficiently light dark 
fermions, a high 
luminosity LHC could provide sensitivity to the invisible decay of the Higgs 
boson.

%%%%%%%%%%%%%%%%%%%%%%%%%
\section{Dark Matter and Dark Photon Phenomenology \label{sec:DM}}
%%%%%%%%%%%%%%%%%%%%%%%%%

The dark fermion sector of our model contains two charged and two neutral Dirac 
fermions $\chi_{1,2}$ and $\xi_{1,2}$.
If the mass of each fermion is less than the sum of the other three 
masses, none of the fermions can decay and all four constitute a stable dark 
matter component.
If one of the fermions has a mass that is larger than the sum of the other 
three masses, it can decay into the lighter three fermions through $W^\prime$ 
exchange. In that case the dark matter will consist of only the lighter three 
fermions.
None of the other massive particles of the model are stable in the regions of 
parameter 
space that we will consider. The heavy dark gauge bosons can decay into a pair 
of dark fermions, while the dark scalar can decay through the Higgs portal into 
a pair of SM particles.
The massless dark photon can have interesting effects in the early universe.

%%%%%%%%%%%%%%%%%%%%%%%%%
\begin{figure}[tb]
\centering
\includegraphics[width=0.2\columnwidth]{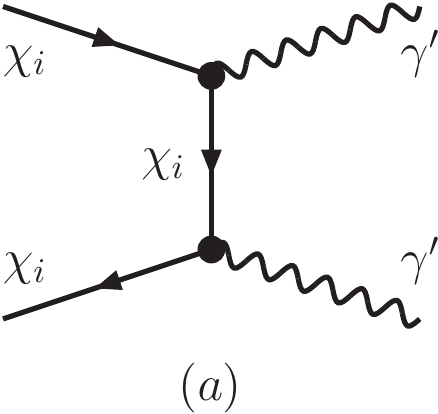}~~~~~~
\includegraphics[width=0.2\columnwidth]{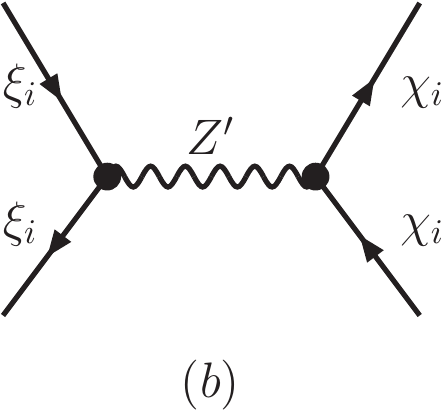}~~~~~~
\includegraphics[width=0.2\columnwidth]{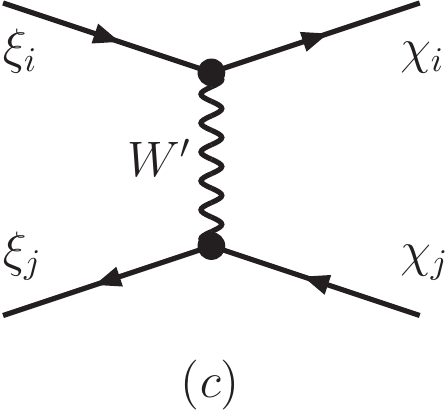}~~~~~~
\includegraphics[width=0.225\columnwidth]{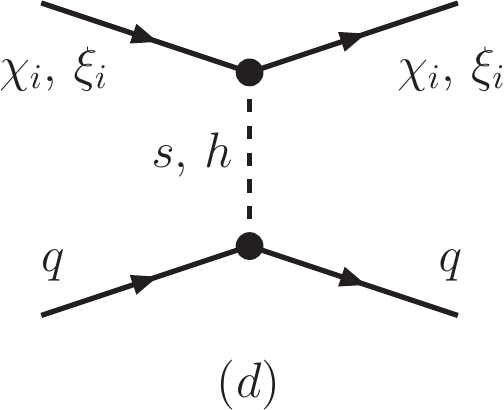}
\caption{Feynman diagrams corresponding to the dominant processes contributing 
to dark matter annihilation (a), (b), and (c), as well as direct detection (d). 
In the case of annihilation into dark photons (a) an additional crossed diagram 
is not shown. \label{fig:diagrams}}
\end{figure}
%%%%%%%%%%%%%%%%%%%%%%%%%

%%%%%%%%%%%%%%%%%%%%%%%%%
\subsection{Dark Matter Relic Abundance}
%%%%%%%%%%%%%%%%%%%%%%%%%

The relic abundance of the charged dark fermions $\chi_i$ is primarily 
set by annihilation into two massless dark photons $\gamma^\prime$. 
Annihilation 
into two dark scalars is
p-wave suppressed and typically negligible. Annihilation into SM particles 
through an s-channel exchange of the dark scalar or the Higgs is strongly 
suppressed 
by the small Higgs portal and therefore also negligible\footnote{Dark matter 
annihilation into SM particles through s-channel exchange of the Higgs or the 
dark scalar might be important in fine tuned corners of parameter space where 
the annihilation is resonant, e.g. $m_{\chi_i} \simeq m_h/2$ or $m_{\chi_i} 
\simeq m_s/2$.}.
For the annihilation cross section into dark photons, depicted in diagram (a) 
of Figure~\ref{fig:diagrams}, we find
\begin{equation} \label{eq:annihilation1}
 (\sigma v)_{\chi_i} \simeq \frac{e_X^4}{8\pi} \frac{1}{m_{\chi_i}^2} ~.
\end{equation}
This annihilation cross section decreases for increasing charged dark fermion 
masses.
The relic abundance of stable charged dark fermions is approximately given by
\begin{equation} \label{eq:relic}
\Omega_{\chi_i} h^2 \simeq 0.12 \times \left( \frac{2.2 \times 10^{-26} 
~cm^3/s}{(\sigma v)_{\chi_i}}\right) ~.
\end{equation}
A charged dark fermion fraction of the total relic abundance is subject to 
various constraints~\cite{Ackerman:mha,Feng:2009mn,Fan:2013yva}.  
A component of (strongly) self-interacting dark matter is constrained by halo 
shapes~\cite{Peter:2012jh} and the observed structure 
of the Bullet Cluster~\cite{Markevitch:2003at,Randall:2007ph}. 
Numerical simulations that account for the observed deviations from spherical 
halos allow for $\sim 10\%$ interacting dark matter, while 
simulations of the Bullet Cluster
allow for up to $\sim 30\%$ of all dark matter to have arbitrarily strong 
self-interactions. A more stringent 
bound comes from possible CMB structure, 
which constrains the fraction of dark matter coupled to dark radiation to 
$\lesssim 
5\%$ \cite{Cyr-Racine:2013fsa}.
If dark 
matter forms a disk due to long ranged interactions, the local dark matter 
density 
puts a comparable bound  on this 
fraction~\cite{Fan:2013yva}. 
In the following, we will therefore allow a charged dark matter fraction of at 
most 5\%.  
This leads to an upper bound on the mass of the charged dark fermions. We find
\begin{equation}
 m_{\chi_1}^2 + m_{\chi_2}^2 \lesssim (1~ \text{TeV})^2 \times e_X^4 ~.
\end{equation}
For values of the dark electromagnetic coupling of the order of the electroweak 
couplings of the SM, $e_X \sim 0.5$, this implies an upper bound on the mass of 
stable charged dark fermions of a few 100 GeV.

It is important to observe, that in the absence of the dark photons, the 
annihilation cross section 
of the fermions $\chi_i$ would be strongly suppressed resulting generically in 
a dark matter relic abundance in excess of the measured value 
$\Omega_\text{DM} h^2 \simeq 0.12$.

Obviously, the neutral dark fermions $\xi_i$ cannot annihilate into the dark 
photons at tree level. Annihilation into two dark scalars or into SM particles 
is also suppressed for the same reasons as in the case of the charged 
dark fermions. The only unsuppressed annihilation of the neutral dark fermions 
is into the charged dark fermions, which is only an option if the neutral 
fermions are significantly heavier than the charged ones, such that their 
freeze out occurs sufficiently earlier. Annihilation into charged dark fermions 
can proceed through s-channel exchange of a $Z^\prime$ or t-channel exchange of 
a $W^\prime$ as shown in diagrams (b) and (c) in Figure~\ref{fig:diagrams}. The 
s-channel exchange of a dark scalar is suppressed by the charged fermion Yukawa 
coupling and hardly relevant in the regions of parameter space that we will 
consider. Even more suppressed is the s-channel annihilation through a Higgs 
boson.
In the limit $m_{\chi_i} \ll m_{\xi_i} \ll m_{Z^\prime}, m_{W^\prime}$, the 
annihilation cross section is approximately given by
\begin{equation} \label{eq:annihilation2a}
 (\sigma v)_{\xi_i} \simeq \frac{m_{\xi_i}^2}{2\pi w^4} \big( 4 s_X^4 - 3 s_X^2 
+ 2 \big) ~.
\end{equation}
We learn that the annihilation cross section increases for increasing neutral 
dark fermion 
mass. 
For $m_{\xi_i} \sim m_{Z^\prime}/2$, the annihilation cross section is strongly 
enhanced by the $Z^\prime$ resonance and reaches its maximum.
For $m_{\xi_i} \gtrsim m_{Z^\prime}/2$ the annihilation cross section decreases 
again with increasing mass. Expressions for the annihilation cross section that 
hold in the general case of arbitrary fermion and gauge boson masses are given 
in the Appendix~\ref{sec:darkmatter}.

%%%%%%%%%%%%%%%%%%%%%%%%%
\begin{figure}[tb]
\centering
\includegraphics[width=\columnwidth]{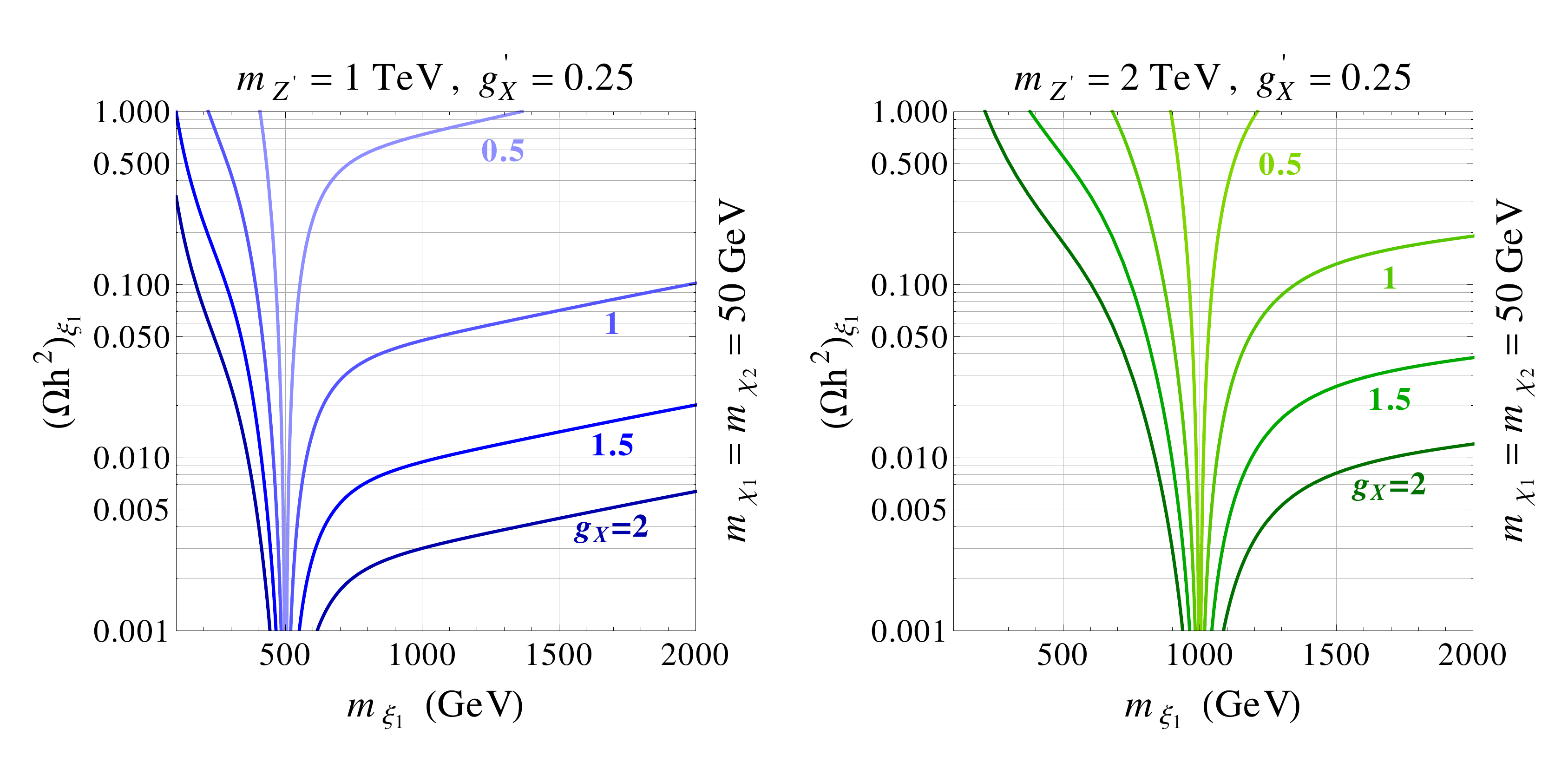}
\caption{The relic density of the light neutral dark fermion species as a 
function of its mass. 
In the left (right) plot, the $Z^\prime$ mass is fixed to 
$m_{Z^\prime} = 1 (2)$~TeV. \label{fig:relic}}
\end{figure}
%%%%%%%%%%%%%%%%%%%%%%%%%

The relic abundance of the stable neutral dark fermions is given by an 
expression analogous to~(\ref{eq:relic}).
The relic abundance of the light neutral dark matter species is shown 
in the plots of 
Figure~\ref{fig:relic} as a function of the fermion mass. In the left and right 
plots the $Z^\prime$ mass is fixed to $m_{Z^\prime} = 1$~TeV and $m_{Z^\prime} 
= 2$~TeV, respectively. The charged fermion masses are fixed to $m_{\chi_1} = 
m_{\chi_2} = 50$~GeV and the dark hypercharge gauge coupling is $g_X^\prime = 
0.25$. The various curves correspond to different choices of the dark $SU(2)_X$ 
gauge coupling that ranges from $g_X = 0.5$ up to $g_X = 2$. We observe that 
the annihilation into charged fermions is very efficient. If the neutral 
fermion is above the $Z^\prime$ resonance, the requirement of the right relic 
abundance leads to an upper bound on the $SU(2)_X$ gauge coupling of the order 
of $g_X \lesssim 1$. Note that a dark matter fermion with mass above the 
$Z^\prime$ mass implies a large fermion Yukawa coupling and therefore 
generically leads to a UV instability in the dark scalar quartic 
$\lambda_\Sigma$. For a dark fermion mass below the $Z^\prime$ resonance as 
preferred by vacuum stability, gauge couplings over a broad range of values can 
be made easily compatible with the 
relic abundance.

%%%%%%%%%%%%%%%%%%%%%%%%%
\subsection{Dark Matter Direct Detection} \label{sec:dd}

The dark matter particles couple to SM particles only through the Higgs 
portal. The direct detection cross section is therefore necessarily suppressed 
by the mixing between the Higgs and the dark scalar. Working with 
scalar mass eigenstates and evaluating the relevant diagram in 
Figure~\ref{fig:diagrams}, we find for the spin-independent cross section for 
elastic scattering of neutral dark matter particles $\xi_i$ off protons
\begin{equation} \label{eq:directdetection}
 \sigma_\text{SI} = \frac{Y_{\xi_i}^2}{2\pi} \frac{m_{\xi_i}^2 m_p^4}{v^2 
(m_{\xi_i} + 
m_p)^2} ~f^2~s_\alpha^2 c_\alpha^2 \left(\frac{1}{m_h^2} - 
\frac{1}{m_s^2}\right)^2 ~,
\end{equation}
where $m_p$ is the proton mass and 
\begin{equation}
 f = \frac{2}{9} + \frac{7}{9}\Big( f_{T_u} + f_{T_d} + f_{T_s} \Big) \simeq 
0.3 
\end{equation}
parametrizes the nuclear matrix element~\cite{Giedt:2009mr}.
The dark matter direct detection cross section is 
suppressed by the 
scalar mixing angle $s_\alpha^2$ as well as by the destructive interference 
between the Higgs and dark scalar exchange. We find typical direct 
detection signals at the level of $\sigma_\text{SI} \simeq 10^{-46}- 
10^{-47}\text{cm}^2$, well below the 
current experimental 
sensitivities of the XENON100 experiment~\cite{Aprile:2012nq} and the LUX 
experiment~\cite{Akerib:2013tjd}. The predicted signals are probably also below 
the sensitivity of XENON1T~\cite{Aprile:2012zx}. They should however be in 
reach of the planned LZ experiment~\cite{Malling:2011va}.

An equation completely analogous to~(\ref{eq:directdetection}) holds also for 
the direct detection cross section of the charged dark matter fermions $\chi_i$.
However, barring additional structure which radically changes the local density 
of the charged dark matter component~\cite{Fan:2013yva}, the maximal relic 
density fraction 
of $5\%$ strongly suppresses sensitivity of  
direct detection experiments to the $\chi_i$. 
For charged dark matter masses of $m_{\chi_i} \sim 50$~GeV, there are only 
corners of parameter space, where the direct detection cross section of 
the charged fermions might reach $\sigma_\text{SI} \sim \text{few} \times 
10^{-48} \text{cm}^2$. Combined with the smaller density of charged dark 
matter, this results in direct detection rates that are at the border of or 
even below the atmospheric and supernova neutrino background, 
and beyond the reach 
of planned direct detection experiments.

%%%%%%%%%%%%%%%%%%%%%%%%%%%%%%%%%%%%%%%%%%%%%%
\subsection{Number of Relativistic Degrees of Freedom in the Early Universe}

The dark photon of our model contributes to the effective number of 
relativistic
degrees of freedom in the early universe. Measurements of the $^4He$ abundance 
\cite{Cyburt:2004yc} in the universe and the combination 
of Planck data with astrophysical measurements of the Hubble constant 
\cite{Ade:2013zuv} put constraints on the active degrees of freedom during Big 
Bang nucleosynthesis 
(BBN) and at the time at which the Cosmic Microwave Background (CMB) radiation 
formed, respectively~\cite{Franca:2013zxa,Fan:2013yva}.

The process of scattering of visible photons into dark photons 
$\gamma \gamma \leftrightarrow \gamma' \gamma'$ can potentially keep 
the dark and the visible sector in thermal equilibrium.
In our model, the Higgs portal is the only connection between the dark 
and the visible sectors.
Therefore $\gamma \gamma \leftrightarrow \gamma' \gamma'$ is induced by 
a dimension eight operator and suppressed by two loops. 
As a consequence, this process decouples at very high temperatures. 
More relevant processes that connect the dark and the visible sector are the 
annihilation of visible photons into dark fermions, 
$\gamma\gamma \leftrightarrow \bar 
\chi \chi$, and of dark photons into SM fermions, 
$\gamma^\prime\gamma^\prime \leftrightarrow \bar 
f f$. Such processes 
are induced by dimension six operators and only suppressed by one loop. 
Depending on the dark scalar mass, mixing 
angle and the dark fermion Yukawa couplings, we find that the decoupling 
temperature is at the order of $T(t_\mathrm{dec})\sim \mathcal{O}(10)$~GeV.   
Below this temperature, the entropy density 
should be 
separately conserved in both sectors, so that the ratio of temperatures 
$\xi(t)= T_\mathrm{dark}/T_\mathrm{vis}$ in the dark and visible sector
at some later time $t$ is given by~\cite{Fan:2013yva}
\begin{align}
 \xi(t)=\left(\frac{g^\mathrm{dark}_{\ast 
s}(t_\mathrm{dec})}{g^\mathrm{vis}_{\ast s}(t_\mathrm{dec})}
\frac{g^\mathrm{vis}_{\ast s}(t)}{g^\mathrm{dark}_{\ast s}
(t)}\right)^{1/3}\,, 
\end{align}
where $g_{\ast s}(t)$ denotes the effective number of degrees 
of freedom at the time $t$.
In the SM, all degrees of freedom besides the Higgs boson, the top, 
and the electroweak gauge bosons are 
active during decoupling, so that  
$g^\mathrm{vis}_{\ast s}(t_\mathrm{dec})=86.75$. In the dark sector, 
the dark photons, and dark fermions can contribute
$g^\mathrm{dark}_{\ast s}(t_\mathrm{dec})=
2+\frac{7}{8}\times 4 \times n=2 + n \times 3.5$,
where $n$ is the number of dark fermions with masses 
below the decoupling temperature. 
At the BBN scale, electrons, neutrinos and photons contribute to the SM 
entropy density 
$g^\mathrm{vis}_{\ast s}(t_\mathrm{BBN})=\frac{7}{8} \times 10 
+ 2= 10.75$, while during formation of the
CMB only colder neutrinos and photons remain active 
$g^\mathrm{vis}_{\ast s}(t_\mathrm{CMB})= 
\left(\frac{4}{11}\right)^{4/3}\times 
\frac{7}{8} \times 6 + 2 = 3.36$. In the dark sector, at these 
times only the dark photon is a relativistic degree of freedom, 
$g^\mathrm{dark}_{\ast s}(t_\mathrm{BBN})=g^\mathrm{dark}_{\ast 
s}(t_\mathrm{CMB}
)=2$. The temperatures in the dark sector during BBN and CMB 
are therefore smaller than in the visible sector. 
We find
\begin{equation}
 \xi(t_\mathrm{BBN})\approx 0.50/0.70/0.82 ~,~~
\xi(t_\mathrm{CMB})\approx 0.34/0.47/0.56~,
\end{equation}
where the first/second/third value corresponds to $n=0/1/2$.
These temperature ratios can be translated into the 
change of effective number of neutrinos at these temperatures
\begin{subequations}
\begin{align}
 \Delta 
N_{\mathrm{eff},\nu}^\mathrm{BBN}&=\frac{8}{7}~\xi(t_\mathrm{BBN})^4 
~\phantom{\left(\frac{4}{11}\right)^{-\frac{
4}{3}}}\approx 
0.07/0.27/0.53~,\\
 \Delta 
N_{\mathrm{eff},\nu}^\mathrm{CMB}&=\left(\frac{4}{11}\right)^{-\frac{
4}{3}}~\frac{8}{7}~\xi(t_\mathrm{CMB})^4 \approx 0.06/0.22/0.43~.
\end{align}
\end{subequations}
In the Standard Model, the effective number of neutrinos is given by 
$N_{\mathrm{eff},\nu}=N_{\mathrm{eff},\nu}^\mathrm{BBN}=N_{\mathrm{eff},\nu}
^\mathrm{CMB}=3.046$ 
\cite{Mangano:2005cc}. 
Currently, the strongest constraints on the numbers of effective degrees of 
freedom during BBN \cite{Cyburt:2004yc}
and 
CMB \cite{Ade:2013zuv} are 
\begin{subequations}
\begin{align}
N_{\mathrm{eff},\nu}^\mathrm{BBN}  &= 3.24^{+0.61}_{-0.57} \quad \text{at 
}\quad 68\%~ \text{C.L.}~,\\
 N_{\mathrm{eff},\nu}^\mathrm{CMB}  &= 3.52^{+0.48}_{-0.45}\quad \text{at 
}\quad 95\%~ \text{C.L.}~.
\end{align}
\end{subequations}
This has to be compared with the values for $ N_{\mathrm{eff},\nu}+\Delta 
N_{\mathrm{eff},\nu}$ in our model, which can be comfortably accommodated 
within the uncertainties.
Future CMB experiments will improve the bounds on 
$N_{\mathrm{eff},\nu}^\mathrm{CMB}$ significantly~\cite{Feng:2014uja} and 
might be able to find evidence for the presence of the dark photon in the early 
universe.

%%%%%%%%%%%%%%%%%%%%%%%%%
\section{Numerical Analysis, Discussion, and Outlook}\label{sec:out}
%%%%%%%%%%%%%%%%%%%%%%%%%

We now analyse the dark scalar and dark matter phenomenology of the model 
numerically, starting from the underlying model parameters in the Lagrangian. 
We explore regions of parameter space that are compatible with vanishing Higgs 
and scalar quartic couplings at the Planck scale.
We checked that small non-zero scalar quartics at the 
Planck scale do not appreciably change any of our findings. 
In addition, we impose the  
correct dark matter relic 
abundance with a $\sim 5\%$ admixture of dark charged dark matter component.

The model introduced in Section~\ref{sec:model} has 9 free parameters: the 
Higgs quartic $\lambda_H$, the dark scalar quartic $\lambda_\Sigma$, the portal 
coupling $\lambda_{H\Sigma}$, the $SU(2)_X \times U(1)_X$ gauge couplings $g_X$ 
and $g_X^\prime$, as well as the Yukawa couplings of the dark fermions 
$Y_{\chi_1}$, $Y_{\chi_2}$, $Y_{\xi_1}$, and $Y_{\xi_2}$. 
We consider the one loop effective scalar potential given in the 
Appendix~\ref{sec:Veff}, with the approximations discussed there. We use 
2 loop beta functions for the SM couplings and 1 loop beta functions for 
the dark sector couplings. For every 
given set of parameters, we minimize the effective potential numerically and 
obtain values for the vevs $v$ and $w$, as well as mass eigenvalues for the 
scalars and their mixing angle. 

We allow to vary the Higgs quartic coupling at the electroweak scale in the 
range $0.259 \lesssim \lambda_H(m_t) \lesssim 0.288$ which is compatible with a 
vanishing $\lambda_H$ at the Planck scale given the current uncertainties on 
the top mass and the strong gauge coupling. 
The portal coupling $\lambda_{\Sigma H}$ sets the ratio of the Higgs vev $v$ 
and the dark vev $w$. For a dark vev at the TeV scale, the portal coupling is 
small, typically at the order of $|\lambda_{\Sigma H}| \sim 10^{-2}$.
We chose a vanishing dark scalar quartic coupling $\lambda_\Sigma$ at the 
Planck scale. The value of the 
scalar quartic coupling at the weak scale as well as the value of its beta 
function are mainly determined by the $SU(2)_X$ gauge coupling $g_X$ and the 
largest dark fermion Yukawa coupling $Y_{\xi_2}$. We chose these parameters 
such that we reproduce the Higgs vev of $v=246$~GeV as well as a Higgs mass of 
$124.5$~GeV $ \lesssim m_h \lesssim 126.5$~GeV. 
With these boundary conditions the heaviest neutral dark fermion turns out to 
be 
unstable.
The lighter neutral dark fermion comprises the dominant part of the dark matter 
relic abundance. Obtaining the right annihilation cross section fixes its 
Yukawa coupling $Y_{\xi_1}$. We allow to vary $Y_{\xi_1}$ such that 
its annihilation cross section lies in the range $1.8 \times 
10^{-26} cm^3/s < (\sigma v)_{\xi_1} < 2.6 \times 10^{-26} cm^3/s$,
reproducing the right dark matter abundance within approximately $20\%$.
For simplicity we assume degeneracy among the charged dark matter particles and 
fix their Yukawa couplings $Y_{\chi_1}$, $Y_{\chi_2}$ such that their masses 
are $m_{\chi_1} = m_{\chi_2} = 50$~GeV. With this mass, requiring that the 
charged dark matter is responsible for $\sim 5\%$ of the observed relic 
abundance fixes the $U(1)_X$ gauge coupling at the TeV scale to be 
at the order of $g_X^\prime \sim 0.25$.
Such small values of $g_X'$ do not significantly impact the running of the dark 
scalar quartic, and therefore cannot overcome the effect of the largest dark 
fermion Yukawa coupling $Y_{\xi_2}$. As a consequence the dark scalar quartic 
will cross zero at the Planck scale.
The limiting case where the scalar quartic 
coupling barely touches zero close to the Planck scale demands a coupling of 
$g_X' 
\sim 0.7$, which results in a charged dark 
matter component well below the percent level. We find that the low energy 
phenomenology of the dark scalar $s$ and the dominant dark matter component 
$\xi_1$ hardly depends on these choices. 

%%%%%%%%%%%%%%%%%%%%%%%
%
\begin{figure}[tb]
\centering
\includegraphics[width=\columnwidth]
{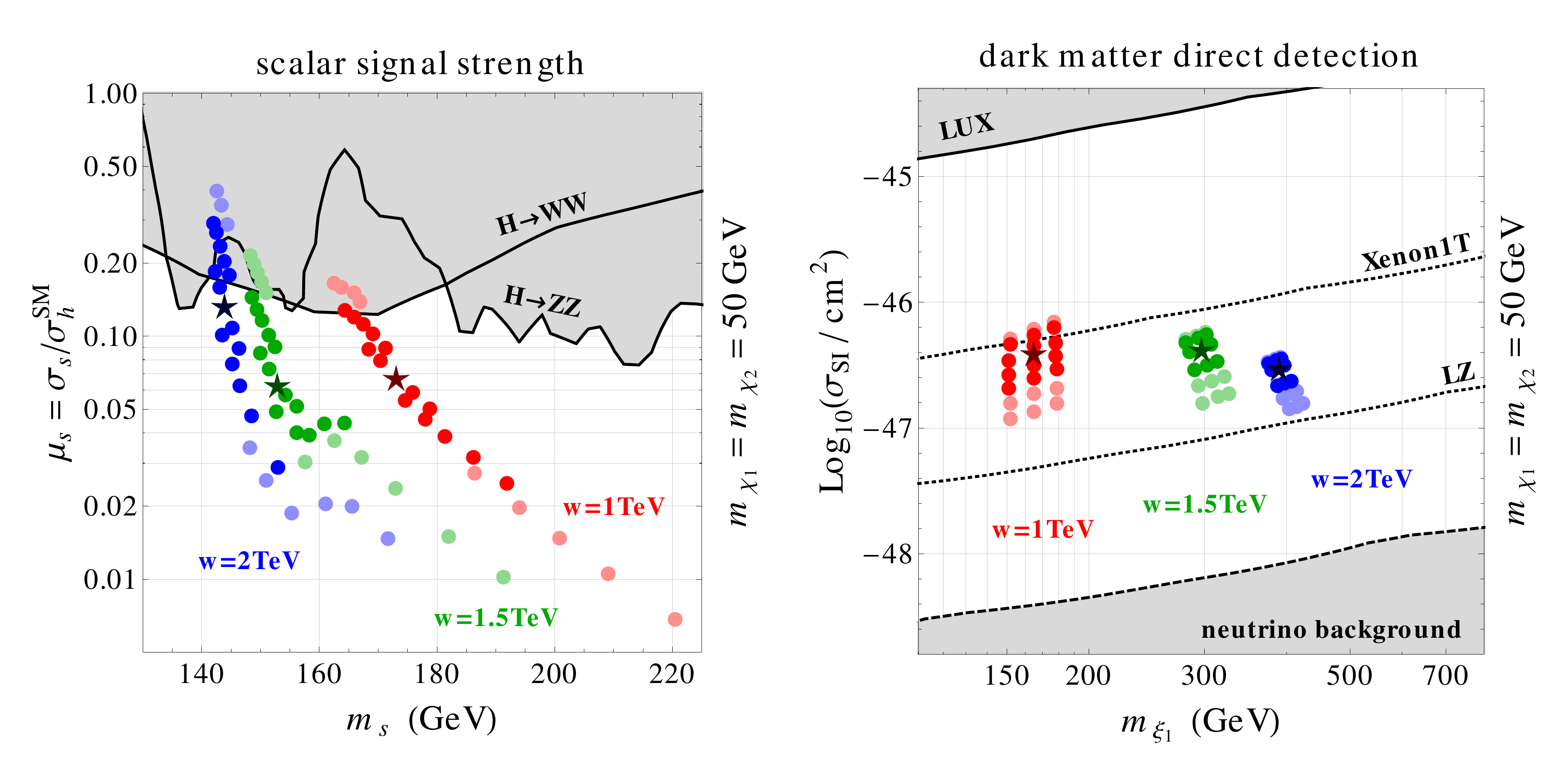}
\caption{Left: Predictions for the signal strength of the dark scalar as 
function of its mass. The shaded regions are excluded by Higgs searches at the 
LHC~\cite{Chatrchyan:2013iaa, Chatrchyan:2013mxa, ATLASC1, ATLASC2}. Right: 
Predictions for the spin independent dark matter nucleon cross 
section. The region above the solid black line is excluded by the current 
experimental bound from LUX~\cite{Akerib:2013tjd}. Future sensitivities of 
XENON1T~\cite{Aprile:2012zx} and LZ~\cite{Malling:2011va} are 
indicated with the dotted lines. In the region below the dashed line, neutrino 
background limits the sensitivity of direct detection experiments.}
\label{fig:predictions}
\end{figure}
%
%%%%%%%%%%%%%%%%%%%

Under the discussed boundary conditions, we obtain predictions for the mass and 
signal strength of the dark scalar, as well as the mass and direct detection 
cross section of the dominant dark matter component. 
In Figure~\ref{fig:predictions}, we show the predictions for direct searches 
for the dark scalar at the LHC (left) as well as dark matter direct detection 
experiments (right). The red/green/blue points correspond to different choices 
for the dark scalar vev $w = 1/1.5/2$~TeV as indicated.
For a fixed choice of $w$, all parameters of the model are determined by the 
chosen boundary conditions discussed above. Demanding that central values 
for $\lambda_H(m_t)$, $m_h$, and, $(\sigma v)_{\xi_1}$ are reproduced,
we obtain the predictions indicated by 
the stars in the plots of Figure~\ref{fig:predictions}. The dark and light 
points show the ranges of predictions that can be obtained by varying the Higgs 
quartic between $0.266 < \lambda_H(m_t) < 0.280$ and 
$0.259 < \lambda_H(m_t) < 0.288$, 
which corresponds to the $1\sigma$ and $2\sigma$ ranges for $m_t$ and 
$\alpha_s$.

We find dark scalar masses in the range 140~GeV~$\lesssim m_s 
\lesssim$~220~GeV. The current sensitivity of Higgs searches in the $ h \to ZZ$ 
and $h \to WW$ channels at the LHC~\cite{Chatrchyan:2013iaa, 
Chatrchyan:2013mxa, ATLASC1, ATLASC2} already starts to probe parts of 
parameter space. Improving the sensitivity down to the percent level over the 
shown mass range, would probe almost the entire parameter space of the 
discussed scenario. Note that in this scenario, the charged dark matter mass is 
fixed to 50~GeV, allowing for a sizable $s \to \chi_i \chi_i$ rate that delutes 
the signal strength at the LHC.
For charged dark matter particles that are heavier than half the dark scalar 
mass, the dark scalar signal strength increases, especially for dark scalar 
masses below the $WW$ threshold, $m_s \lesssim 160$~GeV. 
As expected from the discussion in Section~\ref{sec:higgs}, the prospects 
for direct detection of the dark scalar at the next run of the LHC are 
excellent.

Typical values for the dark matter mass are at the level of few 100~GeV. As 
anticipated already in Section~\ref{sec:dd}, the predicted direct detection 
cross sections are still 1-2 orders of magnitude below the current best 
experimental sensitivity of the LUX experiment~\cite{Akerib:2013tjd}. The 
XENON1T experiment~\cite{Aprile:2012zx} might start to probe parts of the 
parameter space, if the dark scalar vev is around $w \simeq 1$~TeV or 
smaller, which 
corresponds to a small dark matter mass of around $m_{\xi_1}\simeq150$~GeV. 
We expect that 
future dark matter direct detection experiments like LZ~\cite{Malling:2011va} 
will be able to detect the dark matter unless the dark vev is far above the TeV 
scale, in which case the direct searches for the dark scalar become more 
powerful.
The direct detection rates of the charged dark matter component is generically 
below the neutrino floor.
Interestingly, for $m_\chi\leq m_h/2$ the Higgs can decay into the light dark 
matter candidate with a sizable branching fraction. Improved measurements of 
the invisible branching ratio of the Higgs can therefore indirectly constrain 
the mass and 
fraction of charged dark matter even if direct detection experiments cannot see 
it. \\

Anomaly cancellation enforces at least two 
generations of charged fermions with dark charges $Q_X=\pm 1$. In the discussed 
model they are both stable. As argued 
in~\cite{Fan:2013yva}, this can result in dark 
bound states, which imply cosmological dynamics radically different from cold 
dark matter. This would provide a new testing ground for 
our model through measurements of the dark matter distribution within the milky 
way, for example by precisely mapping the movement of stars as planned by the 
GAIA survey~\cite{Famaey}. 
Numerical 
simulations of galaxy formation are beyond the scope of this work, but we 
observe that a double disk scenario as discussed in~\cite{Fan:2013yva} can be 
reproduced within the parameter space of our model. \\

If all physics below the Planck scale should be captured in a model without new 
mass scales, the matter-antimatter asymmetry needs to be addressed. One possibility would be to consider  
baryogenesis at the electroweak scale. This
requires a 
strongly first order phase transition~\cite{Trodden:1998ym,Morrissey:2012db}. 
Since the electroweak scale is 
induced by a dynamically broken dark gauge symmetry, we expect bubble nucleation to occur 
through a two step process in the discussed model. A similar scenario has been 
studied by the authors 
of~\cite{Patel:2012pi} in an extension of the SM with an additional electroweak 
scalar triplet. The sphaleron rate for the dark gauge group 
differs from the one in the SM. A direct comparison to the results 
of~\cite{Patel:2012pi} is therefore not straight forward 
and we leave this interesting question for future 
work.

%%%%%%%%%%%%%%%%%%%%%%%%%
\section{Conclusion \label{sec:con}}

After the 8 TeV run of the LHC, the dynamics of the electroweak symmetry 
breaking mechanism
is still a mystery. 
Natural UV completions of the Standard Model predict new degrees of freedom 
in the vicinity of the 
electroweak scale, that have not been discovered so far. If the 
electroweak scale emerges as a quantum effect from a boundary condition of 
vanishing mass parameters at the Planck scale, the large disparity of scales 
can be explained by RGE running similar to the case of the QCD scale. 
Unlike QCD however, 
there is no underlying symmetry which protects such a boundary condition at the 
Planck scale in the SM.

We have 
shown that clasically scale 
invariant models with a minimal dark sector, that
incorporate radiative electroweak symmetry breaking through 
a 
Higgs portal, and stabilize the Higgs potential up to the 
Planck scale put an upper bound of the order 
of a few TeV on the mass of the dark matter candidate. As discussed previously 
in the literature,
if the dark sector consists of only a scalar charged under dark gauge 
interactions, 
the dark gauge bosons that obtain their mass through 
couplings to the scalar can constitute dark matter.
As the dark gauge coupling has to be sizable in order to generate 
Coleman-Weinberg symmetry breaking, in those scenarios one finds a lower bound 
on 
the dark matter mass. As a consequence, dark matter is generically constrained 
to a window of a few hundred 
GeV to a few TeV. 

In this work we have considered the effect of
additional fermions in 
the dark sector. Radiative symmetry breaking dictates that the bosonic 
contribution to the effective potential needs to be larger than the fermionic 
contribution. This renders the fermions naturally lighter than the gauge bosons 
and allows for
fermionic dark matter masses at the 
electroweak scale, or even below. 
We demonstrated this on the basis of a model that contains a dark sector with a 
$SU(2)\times U(1)$ gauge 
group, a dark scalar that is a doublet under the dark $SU(2)$, and two 
generations of chiral dark fermions. 
The gauge interactions drive the dark scalar quartic negative at low energies 
and radiatively induce a vev for the dark scalar.
The dark sector gauge symmetry is broken spontaneously by the vev of the scalar 
doublet, $SU(2)\times U(1)\rightarrow 
U(1)$, leaving a long-ranged ``dark electromagnetism'' at low energies.
The vev of the scalar doublet also generates a Higgs mass term through a 
quartic portal coupling and thus triggers breaking of the electroweak symmetry 
in the visible sector.

If this is indeed the origin of electroweak symmetry breaking, 
the dark scalar mixes with the Higgs and therefore has Higgs-like couplings to 
Standard Model particles, only suppressed by the portal coupling.
We find that if the dark scalar stabilizes the vacuum up to the Planck scale, 
its mass is 
constrained to be $m_s\lesssim 250$ GeV. 
Its signal strength is generically at the level of $\mathcal{O}(10\%)$ of a SM 
Higgs 
boson. Current Higgs searches in the $WW$ and $ZZ$ channels already constrain 
parts of the parameter space of the model, and the prospects for detecting the 
dark scalar at the next run of the LHC are excellent.
Mixing of the dark scalar with the Higgs also leads to a slight reduction of 
the signal strenghts of the Higgs boson and more precise measurements of the 
various Higgs signal strengths are equally important to test the discussed 
framework.
For sizable mixing, the Higgs boson can also decay through 
the scalar portal into the dark charged fermions, if they are kinematically 
accessible. This can induce an invisible branching ratio of the Higgs of up to 
$\sim 10\%$ which can be within reach of the high-luminosity LHC.

The model has two dark matter components: (i) dark fermions that are charged 
under the long range dark electromagnetism and with masses
typically below the electroweak scale, $m_{\chi_i} \lesssim 100$~GeV; (ii) a 
neutral dark fermion with mass generically in the range $m_{\xi_1} \sim 
100-500$~GeV. 
The model can easily accommodate a non-negligible fraction 
of long-range interacting dark matter of the order of a few percent and could 
have 
interesting implications for galaxy structure formations. While the 
neutral dark matter component has a spin-independent scattering cross section 
with nuclei in reach of future direct detection experiments like XENON1T or 
LZ, the light dark 
matter component will be most likely buried in the 
neutrino background. In addition, the dark 
radiation present in the model can be independently tested by future 
measurements of the number of relativistic degrees 
of freedom in the early universe.
Interestingly, the parameter space of our model in which direct detection experiments are least 
sensitive is the one most strongly constrained by collider searches for the 
dark scalar. The complementarity of both searches imply excellent prospects to discover or 
exclude our model in the near future.

% We have studied the exciting possibility of radiatively generated electroweak 
% symmetry breaking with a dark sector, in which dark fermions naturally 
% constitute dark matter. Near future LHC and direct detection experiments have 
% the capability to test these ideas. In addition

Finally, we argue, that if the electroweak 
scale is generated subsequently 
to the breaking of a dark gauge symmetry, bubble nucleation during the dark and 
electroweak phase transition becomes a two step 
process. Previous studies of a similar scenario suggest, that a strong 
first order phase transition as required by electroweak baryogenesis can be 
achieved in this setup. Studies in this direction are left for future work.

%%%%%%%%%%%%%%%%%%%%%%%%%
\begin{acknowledgments}
We acknowledge helpful discussions with Prateek Agrawal, 
Gia Dvali, Steve Giddings, Chris Hill, Pedro Schwaller, Jessie Shelton and
Alessandro Strumia.
MB and JL acknowledge the hospitality and support of the Theoretical Physics 
Group at SLAC. MC acknowledges the hospitality of MITP.
WA, MC and JL acknowledge the hospitality of the Aspen Center for Physics and 
partial 
support by the National Science Foundation Grant No. PHYS-1066293.
Fermilab is operated by Fermi Research Alliance, LLC, under contract 
DE-AC02-07CH11359 with the United States Department of Energy. The research of 
WA was supported by the John Templeton Foundation. Research at Perimeter 
Institute is supported by the Government of Canada through Industry Canada and 
by the Province of Ontario through the Ministry of Economic Development \& 
Innovation. MB acknowledges the support of the Alexander von Humboldt 
Foundation.
\end{acknowledgments}
%%%%%%%%%%%%%%%%%%%%%%%%%%

%%%%%%%%%%%%%%%%%%%%%%%%%%
%$$\bigstar ~ \bigstar ~ \bigstar ~ \bigstar \hspace{-48.5pt} 
%\raisebox{2pt}{\textcolor{white}{\text{\fontsize{2}{3}\selectfont 54}} 
%\hspace{4pt}
%\textcolor{white}{\text{\fontsize{2}{3}\selectfont 74}} \hspace{4pt}
%\textcolor{white}{\text{\fontsize{2}{3}\selectfont 90}}  \hspace{4pt}
%\textcolor{white}{\text{\fontsize{2}{3}\selectfont 14}}} $$
%%%%%%%%%%%%%%%%%%%%%%%%%%

\begin{appendix}

%%%%%%%%%%%%%%%%%%%%%%%%%
\section{Effective Potential \label{sec:Veff}}

The one loop effective potential $V_\text{eff}$ of our model is approximately 
given by
\begin{eqnarray}\label{eq:Veff_app}
 V_\text{eff}(h,s) &\simeq& \frac{1}{8} \lambda_H(\mu_h) h^4 + \frac{1}{4} 
\lambda_{\Sigma H}(\mu_{sh}) h^2 s^2 + \frac{1}{8} \lambda_\Sigma(\mu_s) s^4 
\nonumber \\
&+& \frac{1}{16\pi^2} \Bigg\{ -3 m_t^2 \left[ 
\log\left(\frac{m_t^2}{\mu_h^2}\right) - \frac{3}{2} \right] \nonumber \\
&& \quad\quad\quad+ \frac{3}{2} m_W^2 \left[ 
\log\left(\frac{m_W^2}{\mu_h^2}\right) - \frac{5}{6} \right] + \frac{3}{4} 
m_Z^2 \left[ \log\left(\frac{m_Z^2}{\mu_h^2}\right) - \frac{5}{6} 
\right]\Bigg\} \nonumber \\
&+& \frac{1}{16\pi^2} \Bigg\{ - \sum_i m_{\chi_i}^2 \left[ 
\log\left(\frac{m_{\chi_i}^2}{\mu_s^2}\right) - \frac{3}{2} \right] - \sum_i 
m_{\xi_i}^2 \left[ \log\left(\frac{m_{\xi_i}^2}{\mu_s^2}\right) - \frac{3}{2} 
\right] \nonumber \\
&& \quad\quad\quad+ \frac{3}{2} m_{W^\prime}^2 \left[ 
\log\left(\frac{m_{W^\prime}^2}{\mu_h^2}\right) - \frac{5}{6} \right] + 
\frac{3}{4} m_{Z^\prime}^2 \left[ 
\log\left(\frac{m_{Z^\prime}^2}{\mu_h^2}\right) - \frac{5}{6} \right]\Bigg\} ~,
\end{eqnarray}
where the field dependent masses are given by
\begin{equation}
 m_t^2 = Y_t^2 h^2 /2 ~,~~
 m_W^2 = g^2 h^2 /4 ~,~~
 m_Z^2 = (g^2 + (g^\prime)^2) h^2 /4 ~, 
\end{equation}
\begin{equation}
 m_{\chi_i}^2 = Y_{\chi_i}^2 s^2/2 ~,~~
 m_{\xi_i}^2 = Y_{\xi_i}^2 s^2/2 ~,~~ 
 m_{W^\prime}^2 = g_X^2 s^2/4 ~,~~ 
 m_{Z^\prime}^2 = (g_X^2 + (g^\prime_X)^2) s^2/4 ~.
\end{equation}
In~(\ref{eq:Veff_app}) we took into account contributions from the top quark, 
the $W$ and $Z$ bosons, the dark fermions and the dark $W'$ and $Z'$ bosons.
Contributions to the effective potential from the Higgs boson $h$, the scalar 
$s$, and the corresponding Goldstone bosons lead to imaginary parts of the one 
loop effective potential, whenever the corresponding quartic coupling 
($\lambda_H$ or $\lambda_\Sigma$) becomes negative. Such imaginary parts signal 
the presence of an instability in the 
potential~\cite{Weinberg:1987vp}.\footnote{The imaginary part coming from the 
Goldstone contribution in the SM is actually spurious and can be avoided by 
resummation~\cite{Martin:2014bca}.} 
We neglect the contributions from $h$, $s$ and the corresponding Goldstone 
bosons. We explicitly checked that this leads to shifts in physical observables 
of a few percent at most. We also do not take into account additional 
corrections coming from the anomalous dimensions of the Higgs and the scalar 
field, as they are typically only at the few percent level, as well. 

All couplings as well as all logarithms in the effective potential depend on a 
renormalization scale. In~(\ref{eq:Veff_app}) we introduced three scales 
$\mu_h$, $\mu_s$, and $\mu_{hs}$ that cancel separately up to terms suppressed 
by two loops. 
In our numerical analysis we set the renormalization scales to the 
corresponding field values $\mu_h = h$, $\mu_s = s$, and $\mu_{hs} = \sqrt{h 
s}$, which is expected to keep higher order corrections to the effective 
potential small.

%%%%%%%%%%%%%%%%%%%%%%%%%
\section{Beta Functions \label{sec:betafunctions}}

The one loop beta functions of the couplings of our framework read ($t = 
\log\mu$)
\begin{eqnarray}
 \frac{d \lambda_H}{dt} = \beta_{\lambda_H} &=&  \beta_{\lambda_H}^{\rm SM} + 
\frac{1}{16\pi^2} 4 \lambda_{\Sigma H}^2  ~, 
\\[8pt]
 \frac{d \lambda_{\Sigma}}{dt} = \beta_{\lambda_\Sigma} &=& \frac{1}{16\pi^2} 
\Big( 12 \lambda_\Sigma^2 + 4 \lambda_{\Sigma H}^2 
- 9 g_X^2 \lambda_\Sigma - 3 (g_X^\prime)^2 \lambda_\Sigma + \frac{9}{4} g_X^4 
+\frac{3}{4} (g_X^\prime)^4 +\frac{3}{2} g_X^2 (g_X^\prime)^2
\nonumber \\ 
&& - 4 \sum_i (Y_{\xi_i}^4 + Y_{\chi_i}^4) + 4 \lambda_\Sigma \sum_i 
(Y_{\xi_i}^2 + Y_{\chi_i}^2) \Big) ~,  \\[8pt]
 \frac{d \lambda_{\Sigma H}}{dt} = \beta_{\lambda_{\Sigma H}} &=& 
\frac{1}{16\pi^2} \Big[ 4 \lambda_{\Sigma H}^2 + 6 (\lambda_H + 
\lambda_\Sigma) \lambda_{\Sigma H} - \frac{\lambda_{\Sigma H}}{2} \Big( 
3 (g^\prime)^2 + 9 g^2 + 9 g_X^2 + 3 (g_X^\prime)^2 \Big) \nonumber \\ 
&& + \lambda_{\Sigma H} \Big( 6 Y_t^2 + 2 \sum_i (Y_{\xi_i}^2 + Y_{\chi_i}^2) 
\Big) \Big] ~,
\end{eqnarray}
\begin{eqnarray}
 \frac{d g_X}{dt} = \beta_{g_X} &=& - \frac{1}{16\pi^2} \frac{39}{6} g_X^3 ~, \\
 \frac{d g_X^\prime}{dt} = \beta_{g_X^\prime}&=& \frac{1}{16\pi^2} \frac{13}{6} 
(g_X^\prime)^3 ~,
\end{eqnarray}
\begin{eqnarray}
 \frac{d Y_{\xi_i}}{dt} = \beta_{Y_{\xi_i}} &=& \frac{1}{16\pi^2} Y_{\xi_i} 
\Big( \frac{3}{2} (Y_{\xi_i}^2 - 
Y_{\chi_i}^2) + \sum_j (Y_{\xi_j}^2 + Y_{\chi_j}^2) - \frac{9}{4} g_X^2 - 
\frac{3}{4} (g_X^\prime)^2 \Big) ~, \\
 \frac{d Y_{\chi_i}}{dt} = \beta_{Y_{\chi_i}} &=& \frac{1}{16\pi^2} Y_{\chi_i} 
\Big( \frac{3}{2} (Y_{\chi_i}^2 - 
Y_{\xi_i}^2) + \sum_j (Y_{\xi_j}^2 + Y_{\chi_j}^2) - \frac{9}{4} g_X^2 - 
\frac{15}{4} (g_X^\prime)^2 \Big)~.
\end{eqnarray}

%%%%%%%%%%%%%%%%%%%%%%%%%
\section{Loop Function \label{sec:loop}}

The loop function that enters the partial width of $h \to \gamma^\prime 
\gamma^\prime$ given in Section~\ref{sec:higgs} reads
\begin{equation}
 f(x) = \left\{ 
\begin{array}{ll} 
\arcsin^2 \sqrt{x} & \quad \text{for} ~x \leq 1 ~,\\
-\frac{1}{4} \left( \log\left( \frac{\sqrt{x} + \sqrt{x-1}}{\sqrt{x} - 
\sqrt{x-1}}\right)  - i \pi\right)^2 & \quad \text{for} ~x > 1~.
\end{array} 
\right.
\end{equation}

%%%%%%%%%%%%%%%%%%%%%%%%%
\section{Dark Matter Annihilation \label{sec:darkmatter}}

In this appendix we give the annihilation cross section of the lightest neutral 
dark fermion $\xi_1$ into the charged dark fermions $\chi_1$, $\chi_2$, that 
are assumed to be lighter than $\xi_1$. Unsuppressed contributions come from 
s-channel exchange of a $Z^\prime$ boson and t-channel exchange of a $W^\prime$ 
boson. We find
\begin{eqnarray}
 (\sigma v)_{\xi_1} &\simeq& \frac{1}{2 \pi} \frac{m_{\xi_1}^2}{w^4} \sqrt{1- 
\frac{m_{\chi_1}^2}{m_{\xi_1}^2}} \left(1 + \frac{m_{\xi_1}^2}{m_{W^\prime}^2} 
- \frac{m_{\chi_1}^2}{m_{W^\prime}^2}\right)^{-2} \nonumber \\
 && + \frac{1}{8 \pi} \frac{m_{\xi_1}^2}{w^4} \sum_{i=1,2} \sqrt{1- 
\frac{m_{\chi_i}^2}{m_{\xi_1}^2}} \left( 1 - 4s_X^2 + 8 s_X^4 + 
\frac{m_{\chi_i}^2}{m_{\xi_1}^2} 2 s_X^2(2s_X^2 -1)\right) \nonumber \\
&& \qquad \times \left( \left(1 - \frac{4 m_{\xi_1}^2}{m_{Z^\prime}^2}\right)^2 
+ \frac{\Gamma_{Z^\prime}^2}{m_{Z^\prime}^2} \right)^{-1} \nonumber \\
&& + \frac{1}{4 \pi} \frac{m_{\xi_1}^2}{w^4} \sqrt{1- 
\frac{m_{\chi_1}^2}{m_{\xi_1}^2}} \left( 1 - \frac{4 
m_{\xi_1}^2}{m_{Z^\prime}^2} \right) \left( 1 - 2 s_X^2 + 
\frac{m_{\chi_1}^2}{m_{\xi_1}^2} s_X^2\right) \nonumber \\
&& \qquad \times \left(1 + \frac{m_{\xi_1}^2}{m_{W^\prime}^2} - 
\frac{m_{\chi_1}^2}{m_{W^\prime}^2}\right)^{-1} \left( \left(1 - \frac{4 
m_{\xi_1}^2}{m_{Z^\prime}^2}\right)^2 + 
\frac{\Gamma_{Z^\prime}^2}{m_{Z^\prime}^2} \right)^{-1} ~.
\end{eqnarray}
The 1st line is the $W^\prime$ contribution, the 2nd and 3rd lines the 
$Z^\prime$ contribution and the 4th and 5th line the interference term. 

The width of the $Z^\prime$ boson that enters the above expressions is given by 
\begin{eqnarray}
 \Gamma_{Z^\prime} &\simeq& \sum_i \frac{g_X^2}{96\pi c_X^2} m_{Z^\prime} 
\sqrt{1 - \frac{4 m_{\xi_i}^2}{m_{Z^\prime}^2}} \left( 1 - 4s_X^2 + 8 s_X^4 - 
\frac{m_{\xi_i}^2}{m_{Z^\prime}^2} \left( 1 + 8 s_X^2 - 16 s_X^4 \right) 
\right)  \nonumber \\
&& ~~ + \sum_k \frac{g_X^2}{96\pi c_X^2} m_{Z^\prime} \sqrt{1 - \frac{4 
m_{\chi_k}^2}{m_{Z^\prime}^2}} \left( 1 - \frac{m_{\chi_k}^2}{m_{Z^\prime}^2}  
\right) ~,
\end{eqnarray}
where the sums over $i$ and $k$ run over those fermions with mass smaller than 
half of the $Z^\prime$ mass.

\end{appendix}

\bibliographystyle{apsrev}
%%%%%%%%%%%%%%%%%%%%%%%%%%%%%%%%%%%%%%%%%%%%%%%%%%%%%%%%%%%%%%%%

\end{document}